\documentclass[aps,prb,twocolumn]{revtex4-1}
\usepackage{graphicx}
\usepackage{float}
\usepackage{times}
\usepackage{color}
\input pdfcolor.tex
\usepackage{graphicx,braket}
\usepackage{epsfig,color}

\begin{document}

\title{The impact of speckle disorder on a superfluid Fermi system}
\author{Abhishek Joshi and Pinaki Majumdar}
\affiliation{Harish-Chandra Research Institute, HBNI, 
Chhatnag Road, Jhusi, Allahabad 211019}
\date{\today}

\begin{abstract}
Optical lattice experiments which probe the effect of disorder on superfluidity 
often use a speckle pattern for generating the disorder. Such speckle disorder
is spatially correlated. While fermionic superfluidity in the presence of 
uncorrelated disorder is well studied, the impact of correlated disorder, 
particularly on thermal properties of the superfluid, is poorly understood. We 
provide a detailed study of the impact of speckle disorder, for varying speckle 
size and disorder magnitude, on the ground state and thermal properties of a 
Fermi superfluid. We work in the coupling regime of BCS-BEC crossover in a two 
dimensional lattice. 
For a fixed disorder strength, an increase in speckle size leads to smoothening 
of the self-consistent background potential, increase in the critical disorder
needed for a superfluid-insulator transition, and an increase in superfluid 
$T_c$. Along with these hints at decrease in effective disorder, speckle
correlations also suppress the superfluid gap and the gap formation temperature
- effects normally associated with increasing disorder.  
We correlate these effects with the effective potential and the single particle
localisation effects in the ground state.
\end{abstract}

\maketitle

\section{Introduction}

Fermi superfluids with $s$-wave symmetry are robust 
to the presence of weak disorder  \cite{anderson,ag}. 
In two dimensions, where all states are 
localised in the presence of arbitrarily weak 
disorder \cite{andloc-2d}, 
and the non interacting system would be an insulator, 
the presence of pairing interaction leads to a superfluid (SF) 
state  \cite{sit-revs}. 
The interplay of disorder and pairing on the 
survival of a superfluid ground state has been extensively explored 
both theoretically \cite{ghosal-prb,triv-1996,dag,dubi,bouadim}
and experimentally 
\cite{haviland-prl,shahar-ovad,escoffier-2004,baturina-2007}.
Most of the earlier experiments are on the solid state, where
multiple interactions may be at play, but artificially
engineered optical lattices \cite{jaks,ess,bloch}
now provide a controlled option.

Weak coupling superfluidity in the presence of disorder 
was first examined by Anderson \cite{anderson}, 
leading to what is called
`Anderson's theorem' about the insensitivity of the SF state
to weak disorder. This insight has been put
on firmer ground by solution \cite{ghosal-prb} 
of the Bogoliubov-de Gennes (BdG)
equations for disordered SF's, explicitly solving 
for the spatially modulated pairing amplitude. 
This leads to significant predictions about cluster formation
and survival of the spectral gap across the disorder driven SF
to insulator transition (SIT). Thermal effects 
can be reasonably accessed within the BdG scheme at weak coupling. 
Complications arise when one moves beyond the weak coupling 
`BCS' window \cite{Meir1,Meir2,tarat1,tarat2,th-int-fermi4}.

For interactions beyond the BCS regime phase fluctuations
of the order parameter, which are neglected in a BdG scheme,
become significant.
Disorder makes the phase stiffness spatially inhomogeneous,
and worsens this situation.  Phase fluctuation 
between weakly coupled clusters
can destroy global coherence with SF order surviving 
in patches. This calls for an approach that treats
thermal fluctuation of the order parameter in an
inhomogeneous situation.  
Full quantum Monte Carlo \cite{triv-1996,bouadim}
accomplishes this, and we have shown earlier 
that a simpler method \cite{tarat1,tarat2} 
can also capture the thermal physics. 
The uncorrelated disorder problem is 
reasonably understood even beyond weak coupling.

It is interesting to ask
how spatial correlations in the disorder - as in disordered
optical lattices - modify the physics.
The `speckle disorder' in these systems is characterised by
two parameters:  the scale $V$ of potential fluctuations, 
and the correlation length, $\sigma$.  Some of the
effects of spatial correlations 
in the disorder have been probed by theory.
For non interacting systems, transport
in a speckle disorder potential 
requires revision of many results that exist in the 
case of uncorrelated disorder.
Several studies have been done on this
\cite{th-non-int1, th-non-int2, th-non-int3, th-non-int4, th-non-int5, 
th-non-int6, th-non-int7, th-non-int8, th-non-int9} 
modifying the Boltzmann
equation and extending the self-consistent theory of localisation.
For interacting
systems we are aware of two kinds of theory, (i)~those which
examine \cite{th-int-bose1,th-int-bose2,th-int-bose3} 
bosonic superfluids in a speckle potential, with 
repulsive interactions present, and (ii)~studies of Fermi
systems \cite{th-int-fermi1,th-int-fermi2,th-int-fermi3}
with repulsive interactions and speckle disorder. 
Those in (i) mainly use the Gross-Pitaevskii framework, 
focusing on the lowest self-consistent eigenstate, while 
(ii)~uses dynamical mean field theory (DMFT).  

\textcolor{red}{
Most of the optical lattice disorder
experiments have been with bosons,
mainly in the `non interacting' regime 
\cite{exp-bose-non-int1, exp-bose-non-int2, 
exp-bose-non-int3, exp-bose-non-int4}
with only a few probing interactions
\cite{exp-bose-int1, exp-bose-int2, exp-bose-int3}.
In the non-interacting problems, the 
trap is switched off
and the Bose gas allowed to expand in the 
presence of speckle disorder. In one dimension (1D), 
even for weak disorder 
\cite{exp-bose-non-int1, exp-bose-non-int2, exp-bose-non-int3}
the cloud stops expanding and forms a 
stationary localised wave.
By fitting the stationary pattern a 
localisation length can be extracted, 
and is found to increase with speckle size.
For bosons in 3D 
\cite{exp-bose-non-int4}.
expansion yields a localised part 
and a diffusive part.
For interacting bosons in a 3D optical lattice
the effect of disorder on the 
condensate fraction has been probed
\cite{exp-bose-int1, exp-bose-int2}.
}

\textcolor{red}{
There are experiments on fermions
probing both the non interacting 
\cite{exp-fermi-non-int0,exp-fermi-non-int1} and interacting 
\cite{exp-fermi-int1,exp-fermi-int2} regimes.
In the non-interacting regime localisation 
has been observed in a 3D 
disordered potential\cite{exp-fermi-non-int0}.
The dependence on 
the correlation length of disorder was 
studied by adjusting the aperture 
of the speckle focusing lens and 
the mean localisation
length was seen to increase linearly 
with speckle correlation length\cite{exp-fermi-non-int1}.
The effect of speckle disorder at fixed 
correlation length was studied 
on a strongly interacting Fermi superfluid \cite{exp-fermi-int2} 
and its properties investigated using high resolution 
{\it in situ} imaging and `conductance' measurements. 
}

\textcolor{red}{
To our knowledge, 
neither theory nor optical lattice experiments have
probed Fermi superfluids with speckle disorder yet.
However, with advances in cold atom technology 
such experiments cannot be far off.}

This paper probes the 
effect of speckle correlated disorder on a Fermi superfluid by 
using the tools we had developed for uncorrelated disorder.
We use the two dimensional attractive 
Hubbard model and work at a mean density
$n=0.9$ per site and intermediate coupling, $U=4t$ 
(where the $T_c$ peaks as a function of coupling).
Our main results are the following.
\begin{enumerate}
\item
{\it Superfluid-insulator boundary at $T=0$:}
The critical disorder for the superfluid 
to insulator transition \cite{footnote} 
increases with speckle size $\sigma$: 
$V_c(\sigma) - V_c(0) \propto \sigma^{\alpha}$, where
we estimate $2.0 \lesssim \alpha \lesssim 2.5$.
\item
{\it Gap and coherence peak:}
At $T=0$,  increasing the disorder at a fixed speckle size
leads to suppression of the coherence peak and the gap.
Increasing the speckle size at fixed disorder 
sharpens the coherence peaks but again {\it suppresses} 
the gap. 
\item
{\it Critical temperature}: At fixed disorder strength 
an increase in speckle size increases $T_c$. We
find that for $V \lesssim V_c(0)$,  $T_c(\sigma) - T_c(0) 
\propto \sigma^{\nu}$, where $\nu \sim 1$. 
Increasing $\sigma$ can convert an insulator
to a superfluid. In such cases, where 
$V > V_c(0)$,  we find $T_c(V, \sigma) \sim 
(\sigma - \sigma_c) \theta(\sigma - \sigma_c)$,
where $\sigma_c$ depends on $V$. 
\item
{\it Thermal pseudogap}: 
While the $T_c$ increases with speckle size, the `gap 
temperature' $T_{g}$ at which the low temperature 
gap converts to a pseudogap, reduces with increasing
speckle size. 
\item
{\it Spatial behaviour and localisation:}
At large speckle size the order parameter $\Delta_i$ 
in the ground state 
is small in the `hill' and `valley' regions 
of the effective potential and is large only over a small
fraction of system volume. The phase stiffness coupling
the $\Delta_i$  is however large due to the delocalisation
promoted by the smooth potential.
\end{enumerate}

\textcolor{red}{
The paper is organised as follows. In Section II we describe
our model and method, including the strategy for generating
the speckle disorder, the Monte Carlo 
method for solving the disordered
Hubbard problem at finite temperature, and the various
indicators in terms of which we characterise the 
disordered superfluid. Sections III-IV are the heart
of the paper. Section III describes our 
ground state results within the Hartree-Fock-Bogoliubov-de 
Gennes (HFBdG) scheme, 
starting with the speckle correlation driven 
smoothening of the order parameter field, the possibility
of an `insulator-superfluid transition', and the unusual
low energy features that emerge in the single particle density
of states.
Section IV is on the finite temperature results, incorporating 
the effect of thermal amplitude and phase fluctuations.
It shows the increase in superfluid $T_c$ and the 
suppression of pseudogap temperature with speckle size.
Section V tries to create an understanding of the results
in terms of the Hartree renormalised effective potential,
the fermionic eigenstates in that potential, and the
effective - spatially inhomogeneous - phase stiffness
that arises in the problem. Many of these require
numerical calculations of their own, but have a simple
connect with localisation theory and XY model physics.
While the results in this section may provide some
insight, the experimentally relevant results are all
in Sections III-IV.
}

\section{Model and method}

\subsection{Model}

We study the attractive Hubbard model in two dimensions (2D),
in the presence of a speckle potential, $V_i$:
$$
H=-t\sum_{<ij>}^{\sigma}c^\dagger_{i\sigma} c_{j \sigma} 
+ \sum\limits_{i\sigma}(V_i -\mu) n_{i\sigma}
-\vert U \vert \sum\limits_{i}n_{i\uparrow}n_{i\downarrow} 
$$
$t$ is the nearest neighbour hopping term, $U$ is the 
onsite attraction,
$\mu$ the chemical potential.  We set $t=1$ 
and fix the fermion density  
at $n \approx 0.9$. 
\textcolor{red}{
We set $U=4t$ where, for the 2D model at the
density we use, the $T_c$ has a peak as a function of $U/t$.
This is the crossover between the weak coupling BCS regime 
and the strong coupling BEC window and the coherence length 
is small enough for our system size $\sim 24 \times 24$.
In the discussion section we show a result on $T_c(U)$.
}

%-----------------------------------------------------------------
\begin{figure}[t]
\centerline{
\includegraphics[width=4.5cm,height=4.5cm]{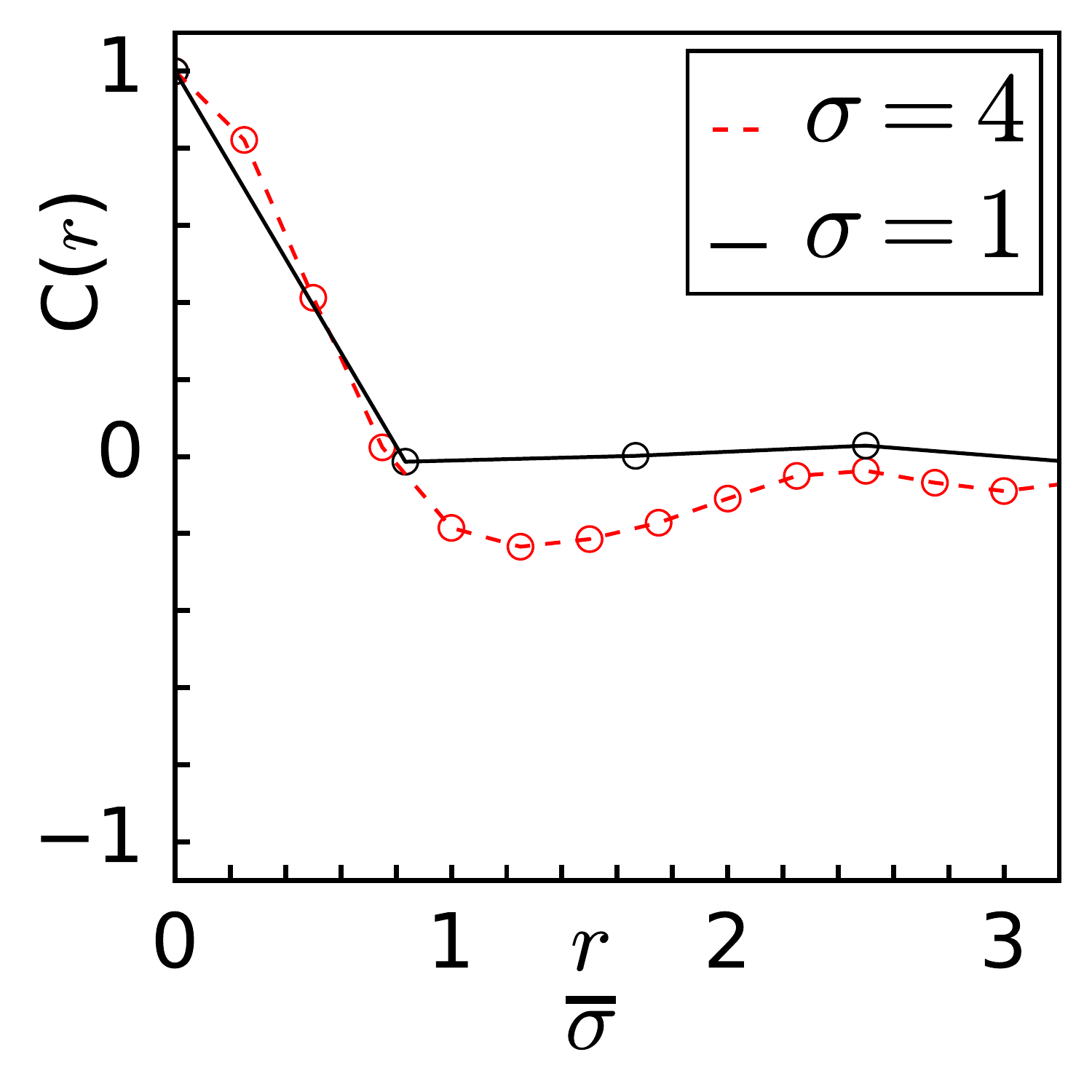} }
\caption{ The speckle correlation function
$(\langle V(\vec{x})V(\vec{x}+\vec{r}) \rangle  -
\langle V(\vec{x}) \rangle^2/\langle V(\vec{x})^2 \rangle) $ for $V =1$,
plotted as a function of $r/\sigma$, by actually sampling our
disorder configurations. The function should be universal, and
 die off for $r/\sigma \gg 1$,
but sampling on a $24 \times 24$ lattice leads to the
non universal features at large $\sigma$.  }
\end{figure}
%-----------------------------------------------------------------

\subsection{Speckle disorder}

\textcolor{red}{
Speckle disorder is spatially correlated. 
We start by generating a spatially uncorrelated 
complex random field, 
$v_i = v_{1,i} + i v_{2,i}$, on the lattice. 
The $v_{a}$ are picked from independent gaussian 
distributions and 
$\langle v_{a,i} \rangle =0$ and
$\langle v_{a,i} v_{b,j} \rangle
= \delta_{ab} \delta_{ij} $,
where $a,~b$ can take value 1 and 2.
The speckle disorder variable is then defined as
\begin{eqnarray}
V_i ~&=&~  V_0 \vert \sum\limits_{\vec{k}}
v(\vec{k})W(\vec{k}) ~e^{i\vec{k}.\vec{r}_i} \vert^2 \cr
v({\vec{k}}) ~&=&~ {1 \over L} \sum\limits_i
v_i e^{-i \vec{k}.{\vec R}_i } \cr
W(\vec{k}) & = &~ 1~~ (k<2\pi/\sigma) \cr
& = &~ 0~~ (k>2\pi/\sigma)
\nonumber
\end{eqnarray}
$V_0 >0 $ and the potential $V_i$ is positive definite. 
It has a distribution $P(V)={1\over V_0}~e^{-V/V_0}$
and a correlation:
$$
\langle V(\vec{x})V(\vec{x}+\vec{r}) \rangle_{\vec x}   =
 V^2(1+\vert\gamma(\vec{r})\vert^2)
$$ 
where $ \gamma(\vec{r})  = 
J_1({\vert\vec{r} \vert}/\sigma)/ ({\vert\vec{r}\vert}/\sigma)
$ with $J_1$  the first order Bessel function and
$\sigma$ the speckle size (correlation length).
Fig.1 shows the two point function generated on
our finite lattice, for two $\sigma$ values.
At large $\sigma$ the long distance behaviour of the 
correlation function deviates from the ideal form on our
$24 \times 24$ lattice.
}

\subsection{Monte Carlo strategy}

\textcolor{red}{
The Hubbard interaction 
cannot be treated exactly so we follow the approach used 
in \cite{tarat1}. 
We write the partition function for the 
model as an imaginary time path integral:
\begin{eqnarray}
Z & = & \int D(\overline{\psi},\psi)e^{-S[\overline{\psi},\psi]} \cr
S & = & 
\int_0^\beta d\tau 
[\sum\limits_{i,\sigma}\overline{\psi}_{i,\sigma}
\partial_{\tau} \psi_{i,\sigma}+ H(\psi, {\bar \psi})]
\nonumber
\end{eqnarray}
The $\psi_i(\tau)$ are Grassmann fields. The quartic term in
the action prevents exact evaluation of $Z$.
To proceed it is usual to introduce auxiliary fields to 
`decouple' the interaction. This Hubbard$-$Stratonovich 
transformation, in terms of (a)~a complex scalar 
`pairing field' $\Delta_i = \vert \Delta_i \vert e^{i \theta_i}$, 
and (b)~a a real scalar field $\phi_i$, is accomplished by
writing:
\begin{eqnarray}
&& exp(U\int d\tau
{\bar \psi}_{i\uparrow}(\tau) {\psi}_{i\uparrow}(\tau) 
{\bar \psi}_{i\downarrow}(\tau) {\psi}_{i\downarrow}(\tau)) \cr
&&~~~~~~~~~~~~~~~~~~~~~~~~~~~~~~~~~~~~~~~~~~~~~~~~~~~~~~~~~~~ 
= \int d {\Delta}_i d{\Delta}^*_i d\phi_i e^{-S_{HS}} \cr
&& S_{HS}  = \int d\tau 
[ \rho_i \phi_i + (\psi_{i \uparrow} \psi_{i \downarrow} \Delta_i + h.c)
+ {1\over U} ( \vert \Delta_i(\tau) \vert^2 + \phi_i^2) 
] 
\nonumber
\end{eqnarray}
The action is now quadratic in the fermions.
Quantum Monte Carlo (QMC) proceeds to sample the fields 
$\Delta_i(\tau), \phi_i(\tau)$, or
their Matsubara components $\Delta_i(i\Omega_n), 
\phi_i(i\Omega_n)$.
We employ a `static path approximation' (SPA) wherein these
fields are assumed to be `$\tau$ independent' (or, alternately,
having only a $\Omega_n =0$ component), but fluctuating
spatially.
The SPA approach retains all the classical amplitude
and phase fluctuations in the model and 
has been elaborately benchmarked in the
BCS-BEC crossover problem.  We show its match 
with QMC in the discussion section.
}

\textcolor{red}{
When the auxiliary fields are assumed to be static, $Z$ becomes:
\begin{eqnarray}
Z ~~~&= & ~\int {\cal D} \Delta {\cal D} \Delta^* {\cal D} \phi
~Tr[e^{- \beta H_{eff}}]  \cr
H_{eff} ~&=& H_0 + H_{coup} + 
{1\over \vert U \vert }
\sum\limits_i (|\Delta_i|^2+\phi_i^2) \cr
H_0~~~ &=& -t\sum_{<ij>}^{\sigma} c^\dagger_{i\sigma}c_{j\sigma}
+\sum\limits_{i\sigma}(V_{i}-{\mu}) n_{i\sigma} \cr
H_{coup} & = & ~~ \sum_{i} (\Delta_ic^\dagger_{i\uparrow}
c^\dagger_{i\downarrow} + h.c) 
+ \sum_i \phi_i n_i \nonumber
\end{eqnarray}
This is a model of quadratic fermions coupled to classical
fields. The Boltzmann weight for the fields can be 
inferred from the top equation:
$ e^{-\beta H(\Delta, \Delta^*, \phi) }
\propto~Tr[e^{-\beta H_{eff}}]$.
The strategy, like in QMC, would be to pick the auxiliary
fields following their Boltzmann weight, solve the 
corresponding fermion problem, and average over
configurations.
}

\textcolor{red}{
Consider the $T=0$ and the finite $T$ cases separately.
(i)~As $T \rightarrow 0$, the system is pushed towards the
maximum probability configuration, {\it i.e}, the minimum of
$ H(\Delta, \Delta^*, \phi)$. The conditions, 
${\partial H}/{\partial \Delta_i} =0$,
${\partial H}/{\partial \phi_i} =0$, {\it etc},
define the usual mean field HFBdG ground 
state.
(ii)~At $T \neq 0$, the fields fluctuate and we generate the
equilibrium $\{\Delta_i,\phi_i\}$ configurations by 
using a Metropolis algorithm. For each attempted update
of the $\Delta_i, \phi_i$ we diagonalise the 
fermionic problem on a $8 \times 8$ 
cluster around the update site and compute the energy
cost of the move \cite{tca}.
}

\subsection{Indicators}

We keep track of the following:
\begin{enumerate}
\item
The ${\bf q}=(0,0)$
component of the pairing field structure factor, 
$$
S({\bf q})={1\over N^2}\sum_{ij}
\langle \Delta^{\ast}_i\Delta_j \rangle
e^{i {\bf q}.(\vec r_i-\vec r_j)}\nonumber
$$
both to determine the presence of superfluidity and also to
locate the $T_c$ scale. Angular brackets indicate thermal
average.
\item
The overall density of states:
$$
~~
 N(\omega) =  
\frac{1}{N} \sum_{n} 
\langle  
|u_{n}|^{2} \delta (\omega - E_{n}) + 
|v_{n}|^{2} \delta (\omega + E_{n})
\rangle
$$
\textcolor{red}{
The $u_n$ and $v_n$ are components of the HFBdG eigenfunctions, and 
$E_n$ are the eigenvalues, in individual equilibrium $(\Delta_i,\phi_i)$
configurations. Since the configurations arise following a Boltzmann weight,
thermal average is same as average over equilibrium configurations.
}
\item
Localisation effects are tracked via the
inverse participation ratio. 
For a normalised state $\vert n \rangle$, the inverse participation
ratio (IPR) is
$P(n) = \sum_i \vert \langle i \vert n \rangle \vert^4$. Averaged over an 
energy interval this leads to:
$$
P(\omega )= {1 \over N(\omega)}
\sum_{n} \delta(\omega - \epsilon_n )P(n) 
$$
$P(\omega )$ is a inverse measure of  the  number of
sites over which eigenstates at energy $\omega$ are spread.
\end{enumerate}

%-----------------------------------------------------------------
\begin{figure*}[t]
\centerline{
\includegraphics[width=12.5cm,height=11.0cm]{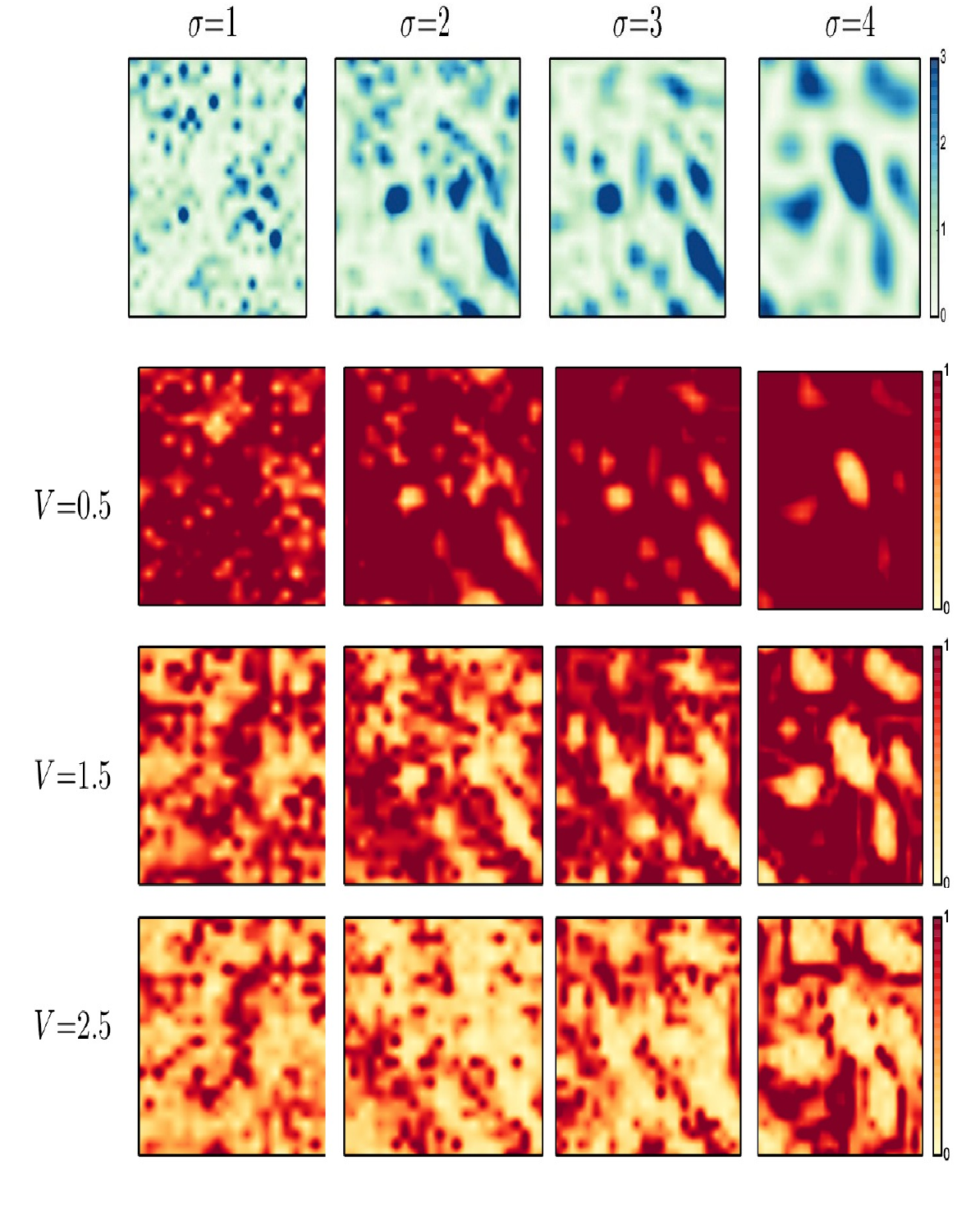}}
\caption{Maps for spatial patterns in the ground state. The top
row shows the disorder potential $V_i$ for fixed $V$ and
four speckle sizes $\sigma$. Patterns for different $V$ can
be generated by simply scaling these up.
Notice the more random $V_i$ landscape at small $\sigma$ and
the progressively smoother variation at larger $\sigma$.
The lower set
of panels shows the amplitude $\vert \Delta_i \vert$ for varying
$V$ and $\sigma$. From the top row down
$V =0.5t,~1.5t,~2.5t$.
The small $V$ large $\sigma$ pattern has the strongest
order while the large $V$ small $\sigma$ panel has the weakest order.
Spatially, $\vert \Delta_i \vert$ anti-correlates with the extremes
in $V_i$.
}
\end{figure*}
%-----------------------------------------------------------------

\section{Ground state}

The ground state of disordered superfluids is characterised by
two spatially varying averages, $\langle n_i \rangle$ and
$\langle c^{\dagger}_{i \uparrow} c^{\dagger}_{i \downarrow} \rangle$.
They are related, in our approach, to the fields $\phi_i$ and
$\Delta_i$. At $T=0$ within our scheme (and mean field theory)
the phase $\theta_i$ is same at every site.
Let us start with the spatial behaviour.

\subsection{Spatial behaviour}

Fig.2 shows the spatial behaviour   in the ground state
for changing speckle size and disorder strength. The top
row shows the pattern of the bare disorder $V_i$ for 
fixed $V$ and four speckle sizes. Realisations with 
larger $V$, but same $\sigma$, can be generated by
simply scaling the potential in the top row. 

As expected, the $V_i$ shows a rapid site to site variation
at $\sigma=1$ and a progressively smoother, island like
structure, at $\sigma =  4$. 
The lower set of panels shows the amplitude, 
$\vert \Delta_i \vert$,
of the pairing field that emerges for different combinations of
$V$ and $\sigma$.
The $V$ values are chosen to capture behaviour at
weak disorder $(V = 0.5t)$, close to critical $(V = 1.5t)$,
and in the insulating regime 
$(V = 2.5t)$ in the uncorrelated limit.
Expectedly, the $\vert \Delta_i \vert$ is large and quasi homogeneous
at small $V$ and large $\sigma$ (top right) and 
rapidly varying and of small average value when $V$ is large
and $\sigma$ is small (bottom left). The $\vert \Delta_i \vert$ 
also `anti correlates' with the extremes $V_i$, since these
regions - with $n_i$ close to $0$ or $2$ - suppress
charge fluctuation. 

While it seems that an increase in $V$ can be `compensated' by an 
increase in $\sigma$, to retain the same degree of overall 
order, the situation is more subtle. 
From Fig.3(a) we located $(V,\sigma)$ pairs where the overall
magnitude of the order is same at $T=0$. 
The pair ($V=1.5t,\sigma=2$) and ($V=2.5t,\sigma=4$) above
satisfy this. 
Spatial similarity? The small $V$ small $\sigma$ pattern,
while inhomogeneous, has a more `space filling' character in 
the order parameter compared to the large $V$ large $\sigma$ 
case where a small fraction of the total 
area has large $\Delta_i$ and large areas have 
$\Delta_i \rightarrow 0$.
The large hills and valleys created by the strong
$V$ large $\sigma$ make 
the $\Delta_i$ pattern more filamentary.
{\it An increase in $\sigma$ is not simply 
like a decrease in $V$.  }

\subsection{Phase diagram}

While the local distribution follows $P(v) \propto e^{-v/V}$,
the introduction of a correlation length
$\sigma$ makes the critical disorder $V_c$ dependent on
$\sigma$. Fig.3(a) shows
how the SF order parameter at $T=0$ (obtained by
extrapolating the finite $T$ result) falls with $V/t$
for different $\sigma$. The intersection of these lines with the
$x$ axis maps out $V_c(\sigma)$.

Fig.3.(b) shows the $V-\sigma$ ground state phase diagram
obtained by the method above. For uncorrelated exponential
disorder we find $V_c \sim 1.8t$. With increase in $\sigma$
the $V_c$ increases - widening the SF window - and we find
that $V_c(\sigma) - V_c(0) \propto \sigma^{\alpha}$ with
$\alpha \sim 2.4$.
We call the phase without SF order an `insulator' since
it has an interaction induced gap in the spectrum.

%-----------------------------------------------------------------
\begin{figure}[t]
\centerline{
\includegraphics[width=4.1cm,height=4.4cm]{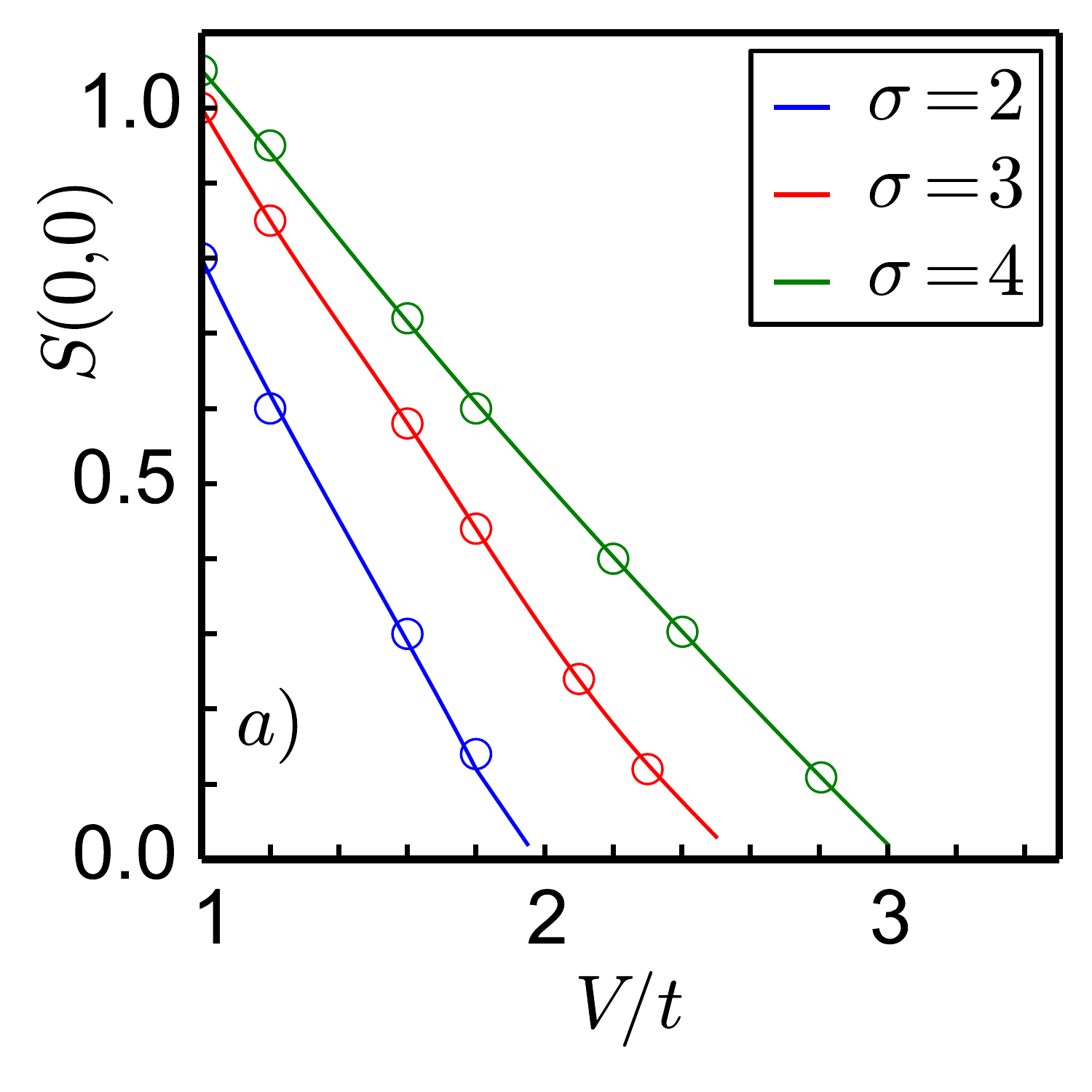}
\includegraphics[width=4.1cm,height=4.4cm]{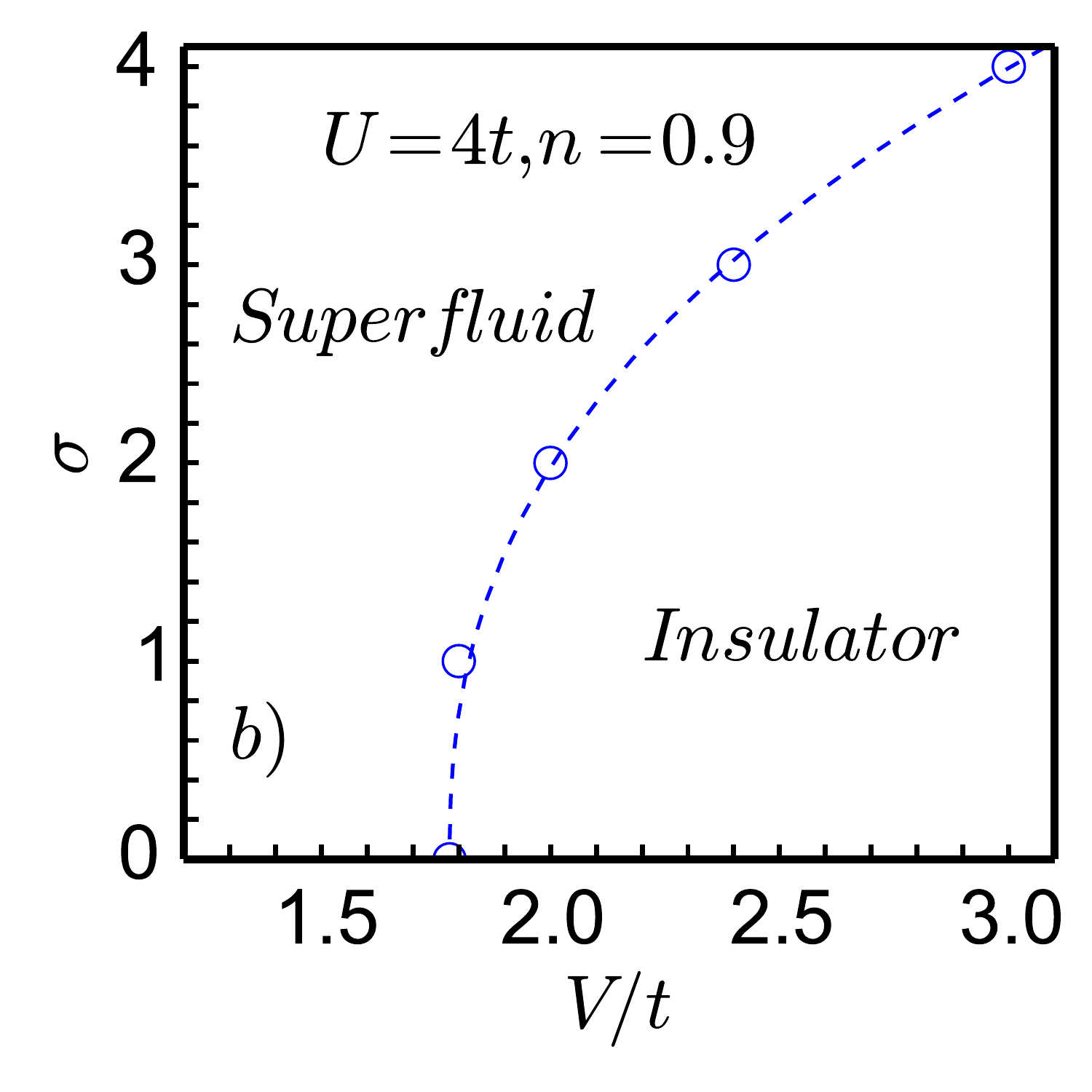} }
\caption{Order in the ground state: (a).~The superfluid order
parameter, {\it i.e}, the pairing field structure factor
$S({\bf q} =(0,0))$, extrapolated down to $T=0$, as a function
of disorder strength $V$ at various speckle size $\sigma$. The
critical disorder for SF to insulator transition increases with
$\sigma$. (b)~The ground state phase diagram at $U=4t$ and $n=0.9$
that emerges from the data in panel (a).  The dotted line is a fit
$V_c(\sigma) - V_c(0) \propto \sigma^{2.4}$.  }
\end{figure}
%-----------------------------------------------------------------

%-----------------------------------------------------------------
\begin{figure}[b]
\centerline{
\includegraphics[width=4.2cm,height=3.0cm]{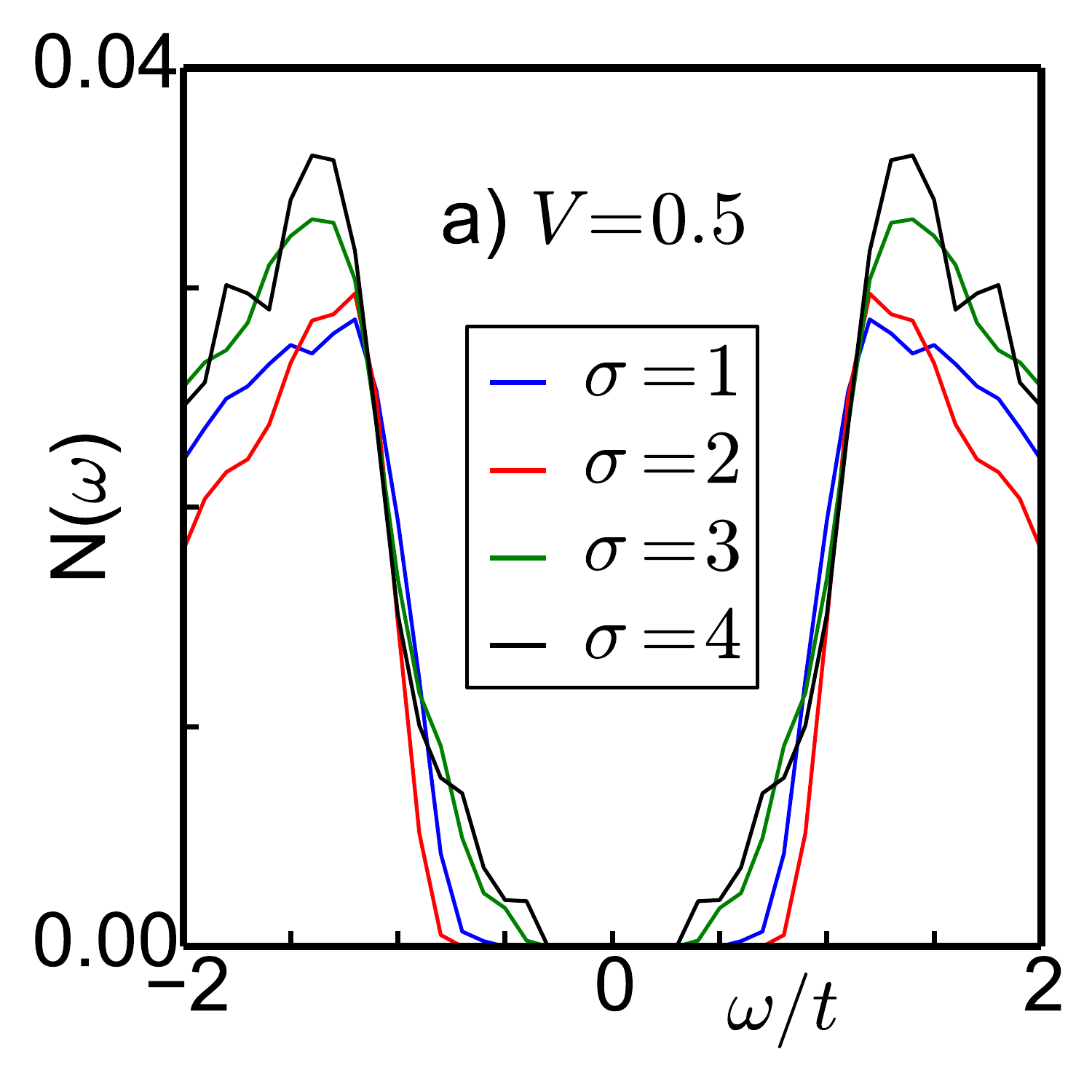}
\includegraphics[width=4.2cm,height=3.0cm]{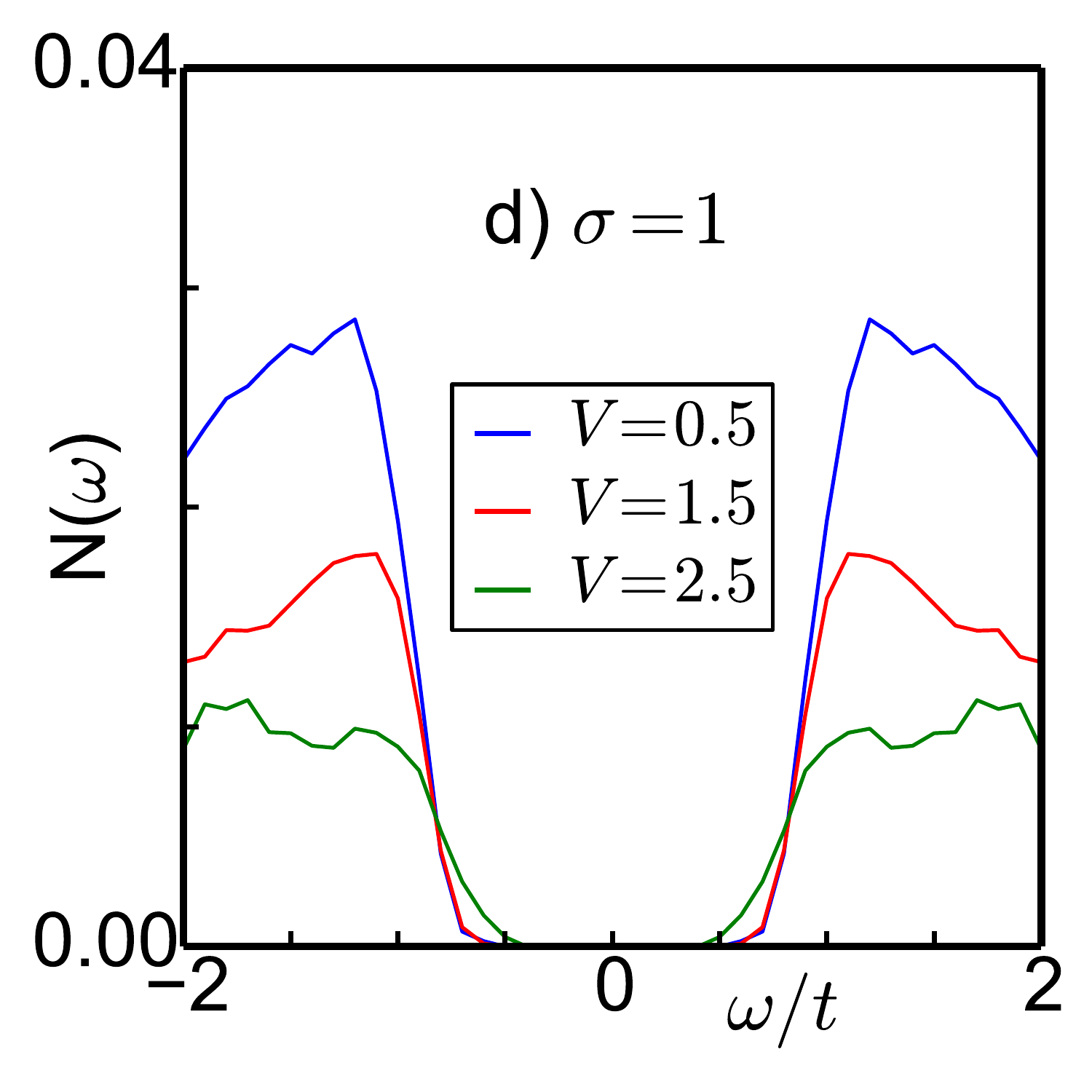} }
\centerline{
\includegraphics[width=4.2cm,height=3.0cm]{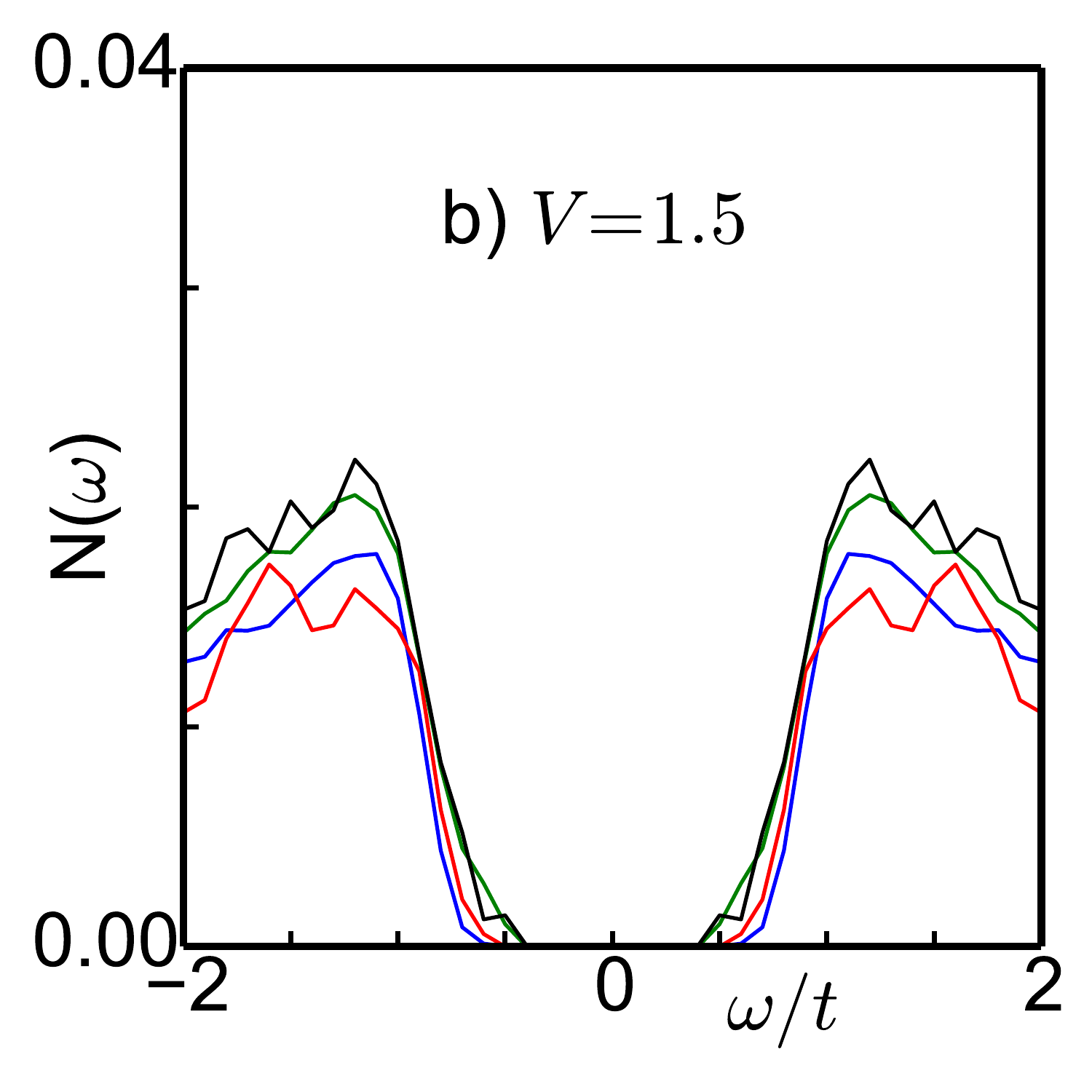}
\includegraphics[width=4.2cm,height=3.0cm]{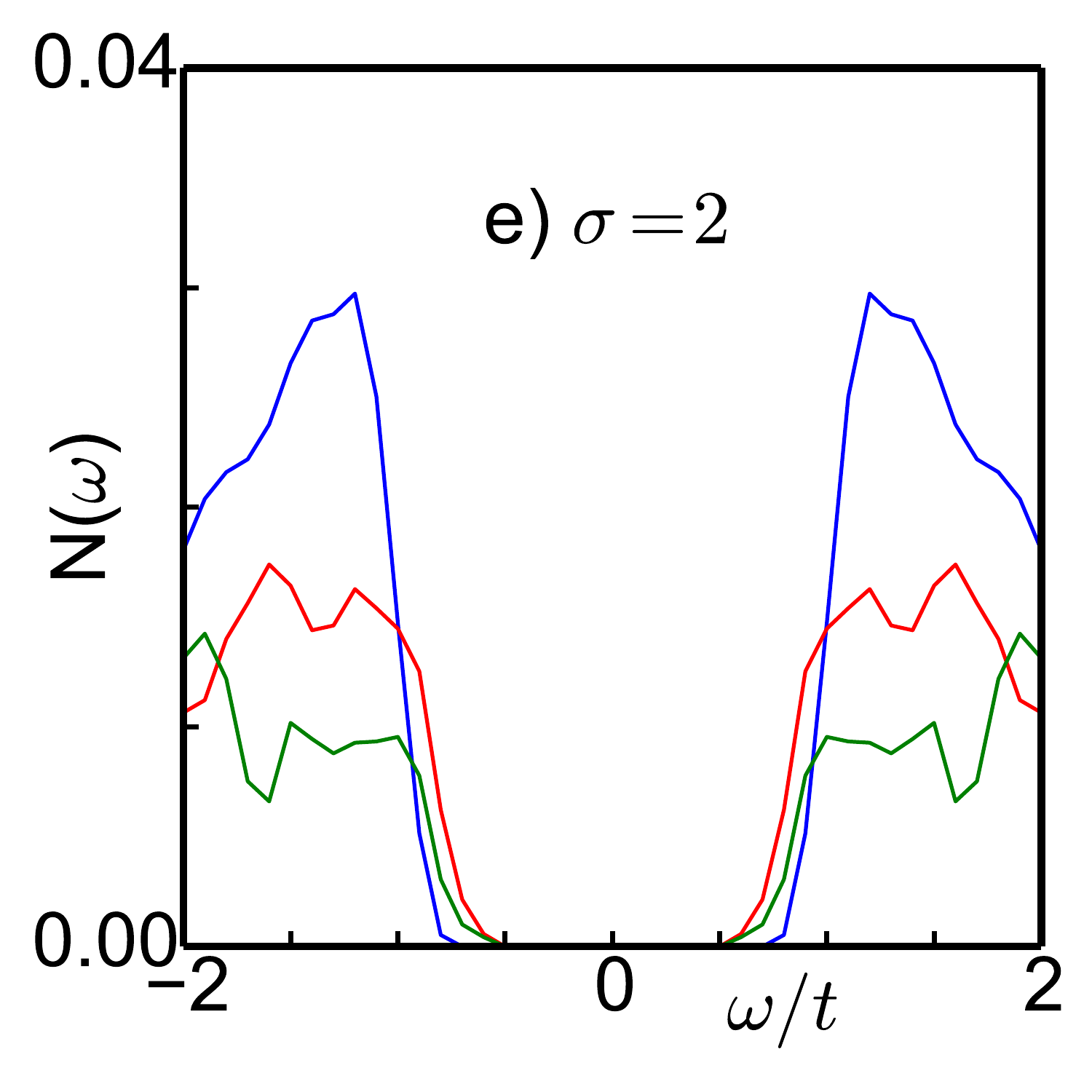} }
\centerline{
\includegraphics[width=4.2cm,height=3.0cm]{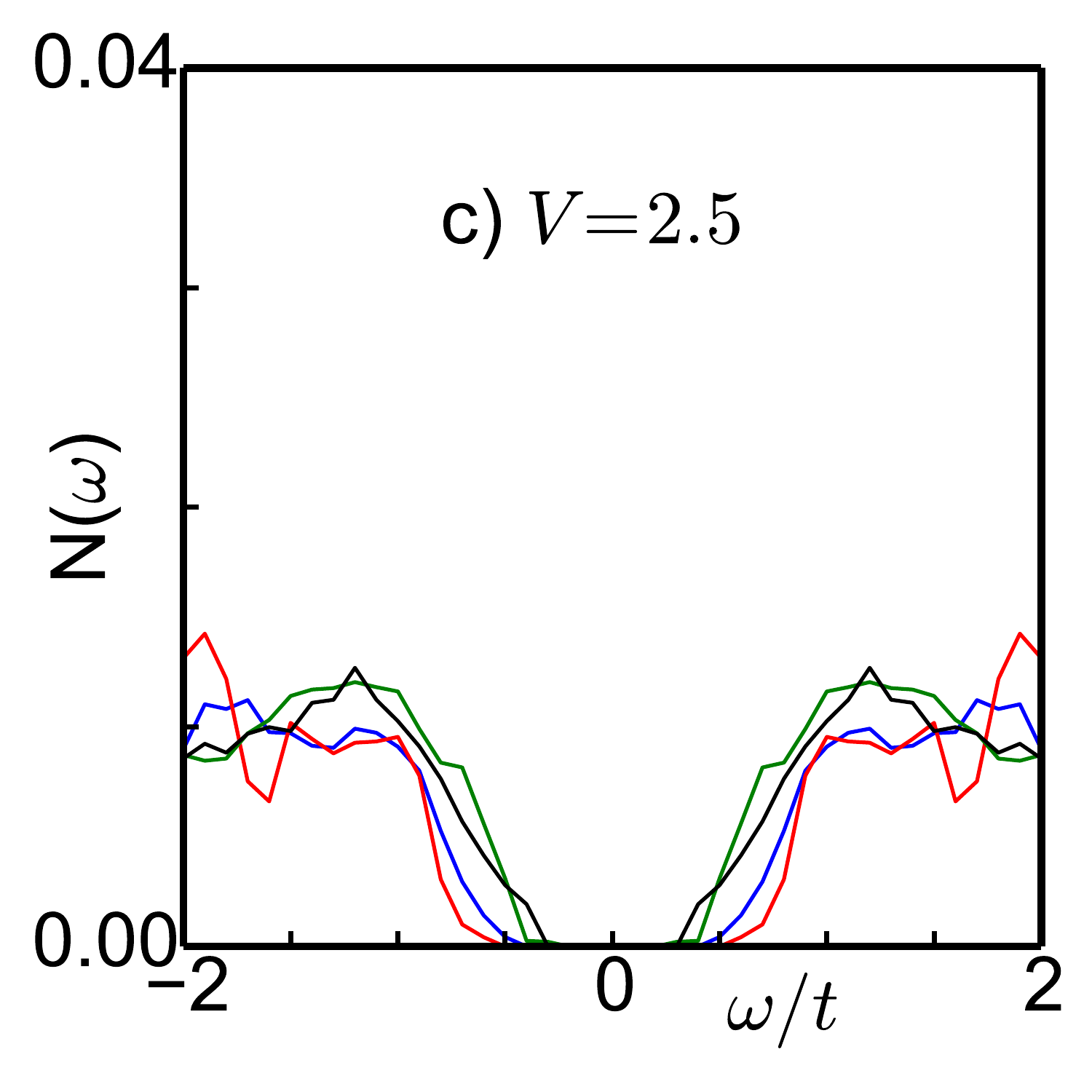}
\includegraphics[width=4.2cm,height=3.0cm]{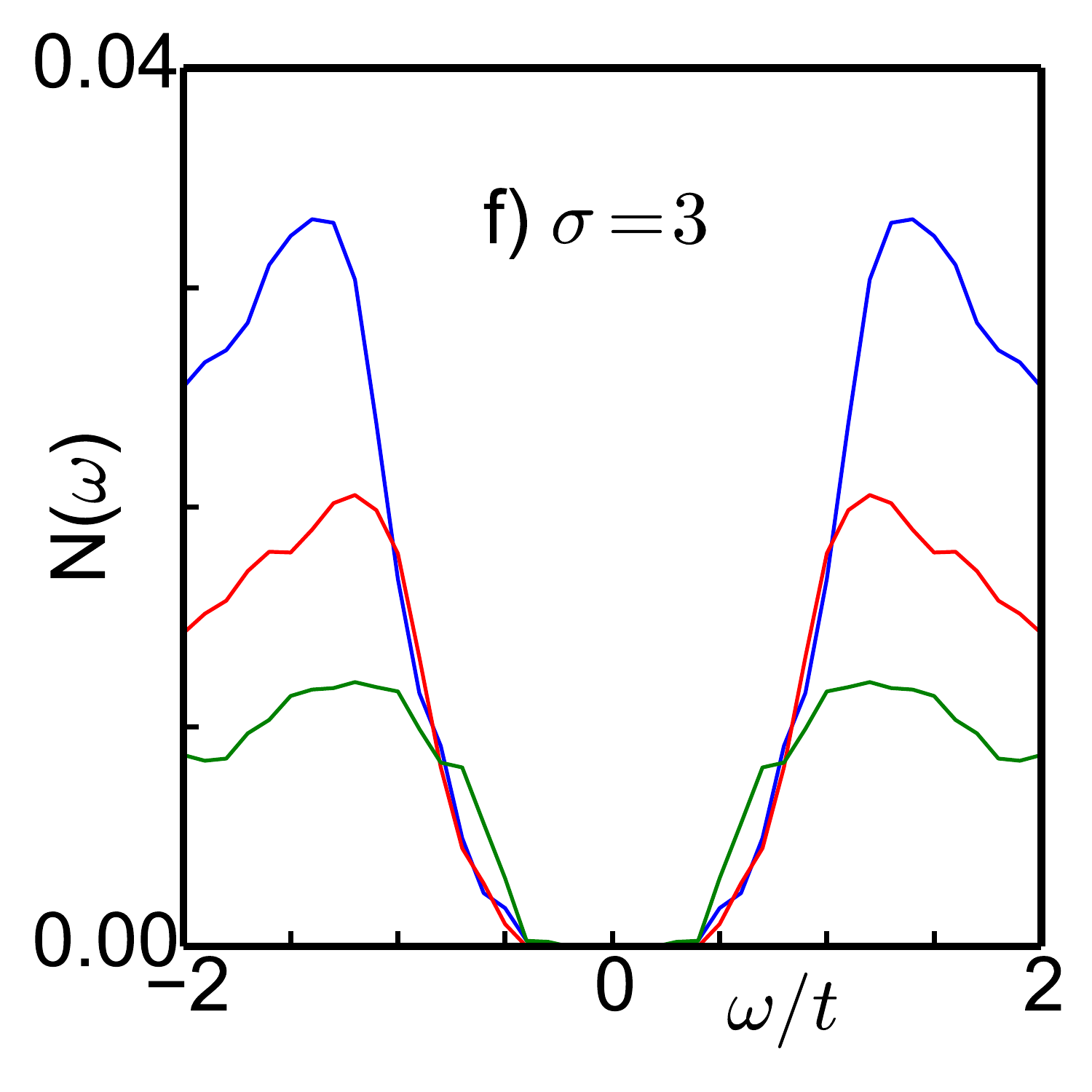} }
\caption{ Density of states in the ground state. Panels (a)-(c) show the
DOS at three strengths of disorder $V$ and for four $\sigma$ at each $V$.
Increase in $\sigma$ sharpens the coherence peak but also suppresses the
gap.  Panels (d)-(f) show the same data as in (a)-(c) now highlighting
the variation with $V$ at fixed $\sigma$. Here increasing $V$ suppresses
both the coherence peak and the gap.  }
\end{figure}
%-----------------------------------------------------------------

\subsection{Density of states}

We now examine the DOS in the ground state for varying $V$ and
$\sigma$, Fig.4. The left column shows results at fixed $V$, while the
right column shows the same data organised in terms of fixed $\sigma$.
Panel (a) shows the $\sigma$ dependence at weak disorder.
There are three effects that emerge on increasing $\sigma$:
(i)~the `coherence peak' sharpens, (ii)~the gap in the DOS
reduces, and (iii)~the rise from the gap edge to the
coherence peak shows a reducing slope - unlike the sharp
rise that one sees in a clean system.  Panels (b) and (c)
show behaviour similar to (a) except
for an overall suppression in magnitude (the bandwidths are
much larger here) and a rather tenuous coherence feature.

%-----------------------------------------------------------------
\begin{figure}[b]
\centerline{
\includegraphics[width=8.2cm,height=4.0cm]{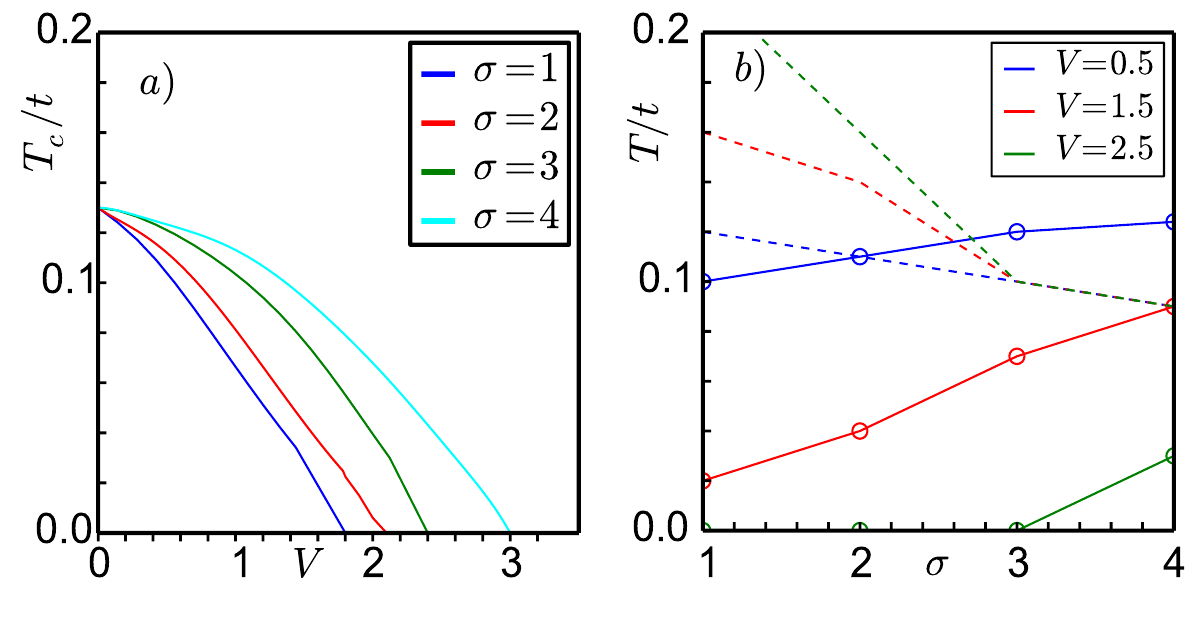}}
\caption{ Variation of the superfluid $T_c$ with disorder and
speckle correlation length. (a)~Disorder dependence for various
$\sigma$. At $V=0$ we have the `clean' $T_c$.  The rate of fall
with $V$ decreases with increasing $\sigma$.  (b)~Firm lines:
$T_c$ for varying $\sigma$ and three values of $V$.  Dotted lines:
$T_g$ - low $T$ gap to high $T$ pseudogap crossover temperature.  }
\end{figure}
%-----------------------------------------------------------------
%-----------------------------------------------------------------
\begin{figure*}[t]
\centerline{
\includegraphics[width=13.7cm,height=13.0cm]{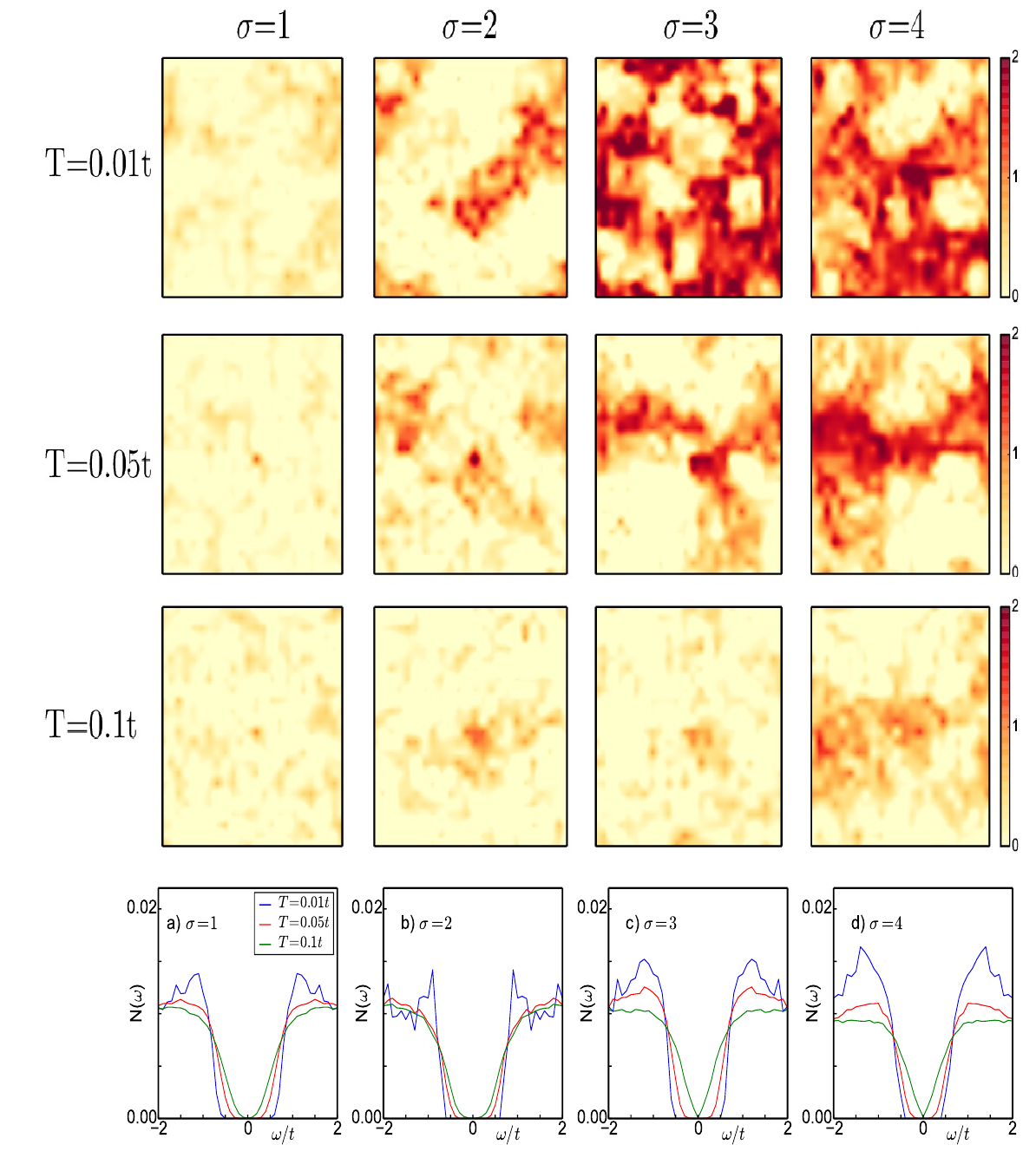}
}
\caption{Speckle size and temperature dependence of $\Delta_0.\Delta_i$
at strong disorder.  $\Delta_0$ is the pairing field at a reference
(corner) site.  The overlap is based on a single MC configuration at
$V=2t$ and the $\sigma, T$ indicated.  Increasing $\sigma$ augments
intersite correlation, with the largest $\sigma$ lowest $T$ panel
(top right) having the strongest correlation.  Bottom row: Density of
states at $V = 2t$, showing $\sigma$ and $T$ dependence.  At low
$\sigma$ the low $T$ state shows no coherence peak, and a broad gap
that smears out with increasing $T$.  At $\sigma =4$ there is a
reasonable coherence feature, and a smaller gap at low $T$. The gap
fills more quickly with rising $T$ than at $\sigma=1$. The behaviour
in panels (b)-(c) interpolate between (a) and (d).
}
\end{figure*}
%-----------------------------------------------------------------

Feature (i) above would be expected also in terms of a decrease
in the effective disorder, (ii) and (iii) however contradict
that interpretation. As we will see, an increase in
$\sigma$ does not make the $\vert \Delta_i \vert$ large and
homogeneous, it makes the $\vert \Delta_i \vert$ distribution
very broad, with large weight at small $\vert \Delta \vert$.
This leads to the low energy weight and
unusual shape in $N(\omega)$.

The disorder dependence at fixed $\sigma$ is more traditional.
Increasing $V$ suppresses the coherence peak and increases
low energy spectral weight - effectively reducing the gap.
The feature is visible in panels (d)-(f).

\section{Thermal fluctuations}

To understand how speckle correlations affect the
$T_c$ and spectral properties
we do a Monte Carlo on $H_{eff}$, annealing the
variables $\phi_i$ and $\Delta_i$, which now pick up a
distribution at each site. We compute spatial correlations
and DOS averaged over equilibrium configurations.

\subsection{Phase diagram}

Tracking the ordering peak, $S({\bf q} =0)$ in the pairing
structure factor allows us to locate a transition scale. For
our density and interaction choice that scale is $\sim 0.13t$
in the clean limit. We choose several $V$ for $\sigma = 1-4$,
and about 10 realisations for each $(V,\sigma)$ combination,
and cool the system from high temperature. We save
equilibrium configurations of $\{\Delta_i,\phi_i\}$,
the pairing structure factor $S({\bf q})$, and the
DOS.

The $T_c$ scale that emerges is shown in Fig.5. Panel (a)
shows the $V$ dependence for different $\sigma$ while
panel (b) shows the $\sigma$ dependence at fixed $V$.
In (a) all the $T_c$ curves start at the clean value
when $V =0$. The drop with  $V/t$ is relatively quick
at $\sigma=1$, hitting $T_c=0$ at $V \sim 1.8t$, while
at $\sigma =4$ the fall is much slower and the critical
disorder is $V \sim 3t$. These numbers lead to the
phase diagram in Fig.3.

Panel (b) shows the effect of increasing $\sigma$ on the
$T_c$, for fixed values of $V$. At weak disorder, $V = 0.5t$,
the $T_c$ rises slowly with increasing $\sigma$ and tends
to reach the clean limit value for $\sigma \gtrsim 4$. At $V = 1.5t$,
where for uncorrelated disorder the system is close to a SF-insulator
transition, increasing $\sigma$ leads to a quicker rise in $T_c$.
The third case,
at $V = 2.5t$ is the most interesting. Here the system remains
insulating upto $\sigma \sim 3$ and only at $\sigma =4$ do we
see a small finite $T_c$. This is a speckle size induced insulator
to SF transition - the bottom right panel in Fig.2 suggests that
this occurs via percolation.
The dotted lines indicate a crossover from the low $T$ `gapped'
regime to a higher $T$ pseudogap regime. The corresponding
$T_g$ scale reduces with increasing $\sigma$.

%-----------------------------------------------------------------
\begin{figure*}[t]
\centerline{
\includegraphics[width=15.2cm,height=9.6cm]{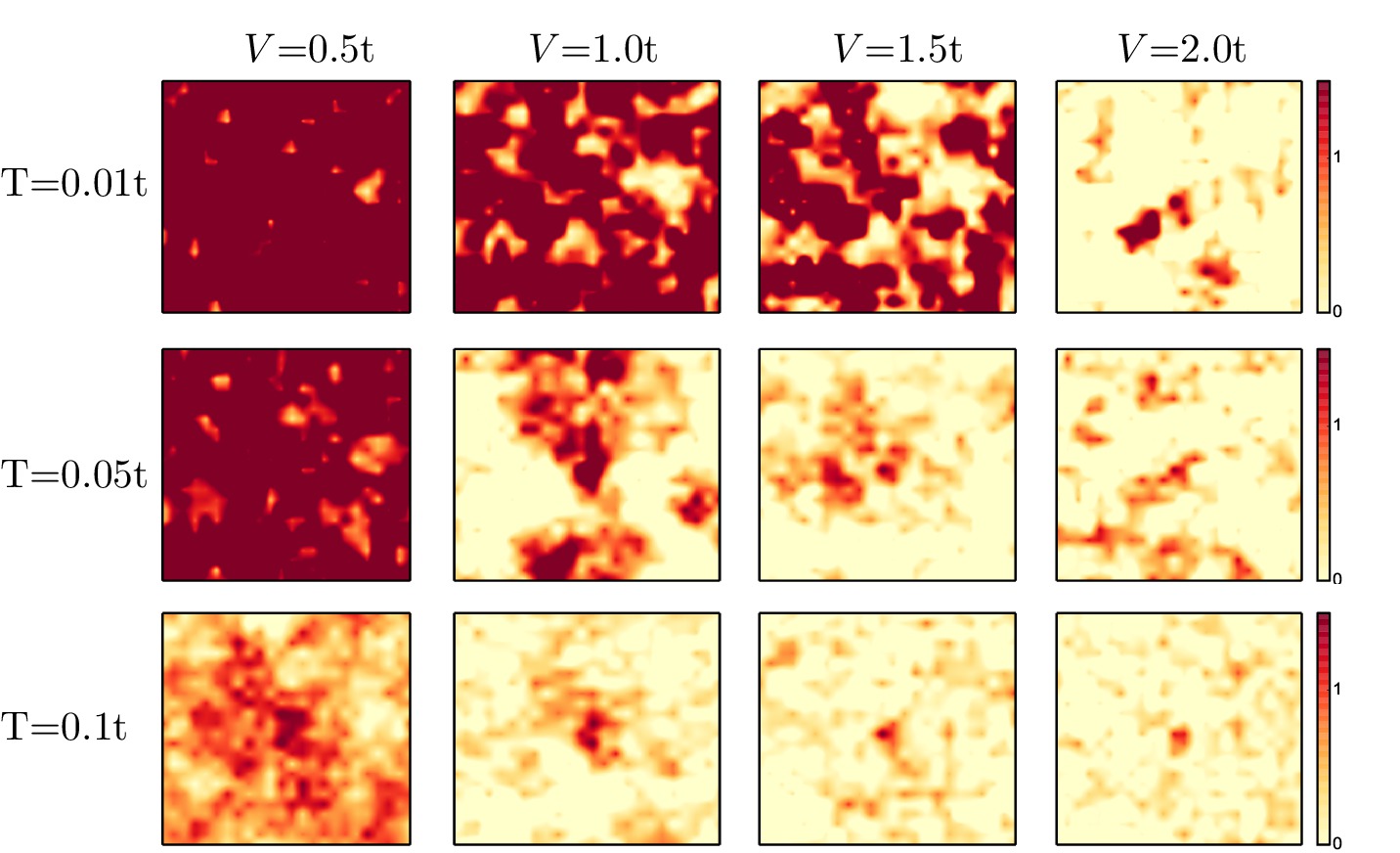}
}
\vspace{.2cm}
\centerline{
~~~
\includegraphics[width=3.20cm,height=3.4cm]{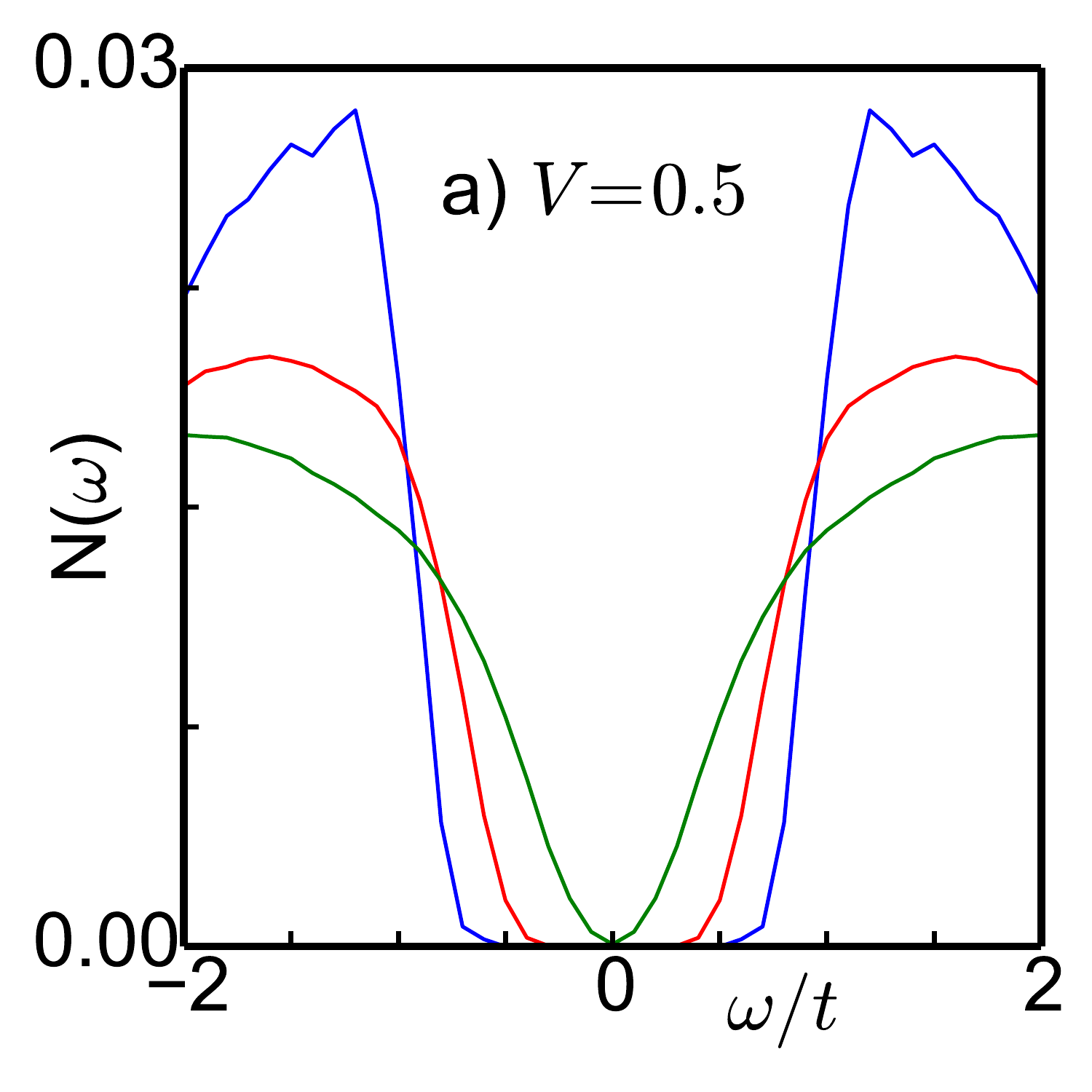}
\includegraphics[width=3.20cm,height=3.4cm]{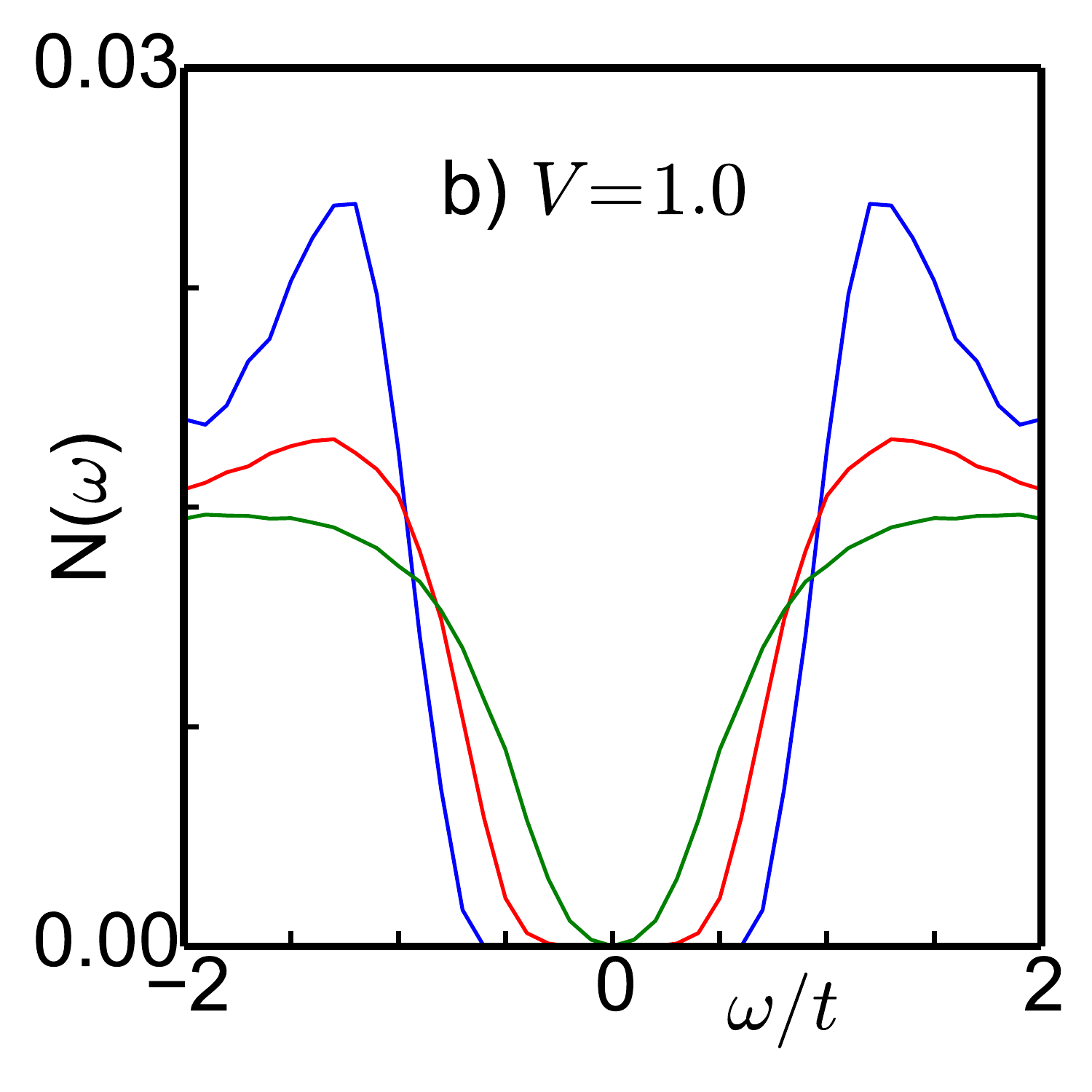}
\includegraphics[width=3.20cm,height=3.4cm]{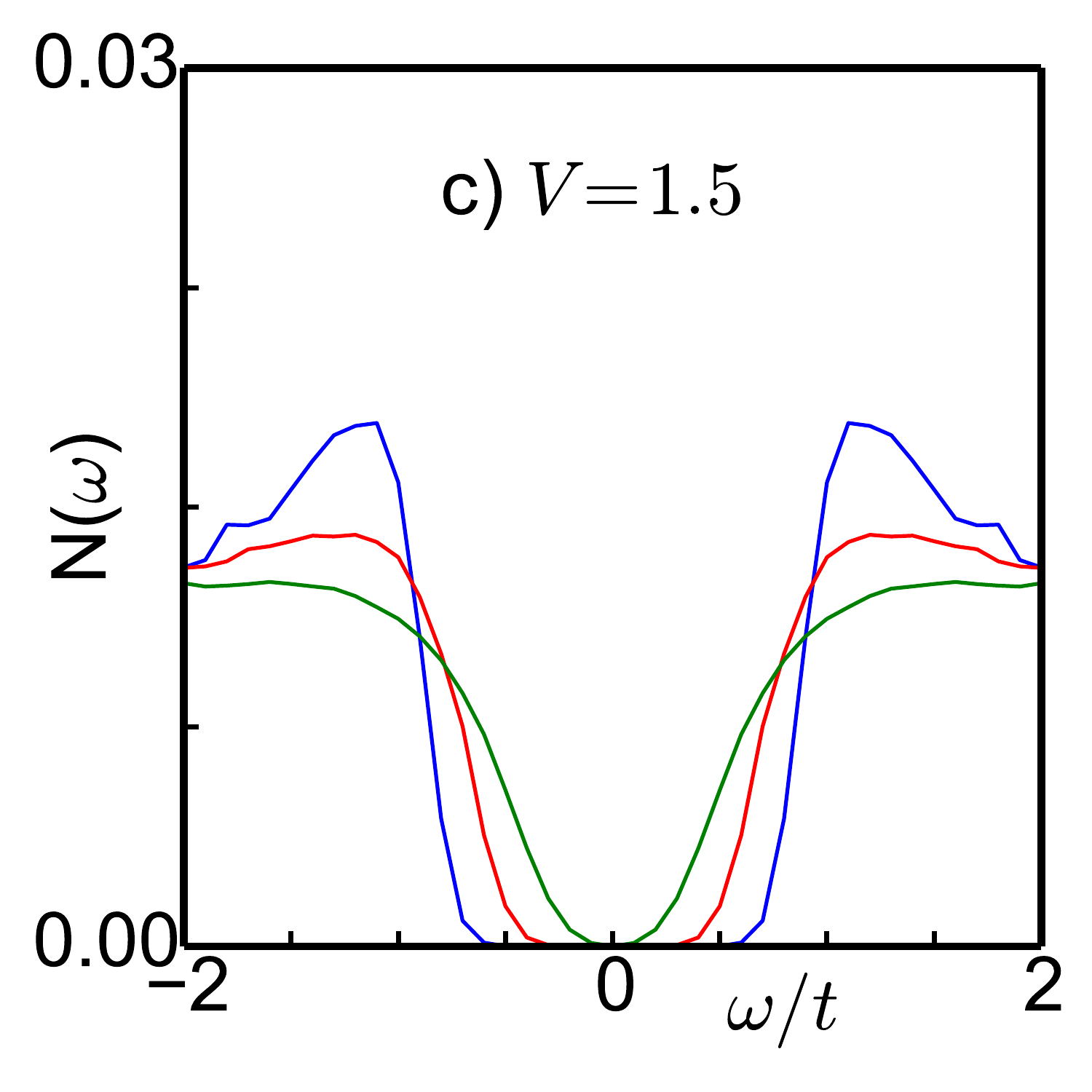}
\includegraphics[width=3.20cm,height=3.4cm]{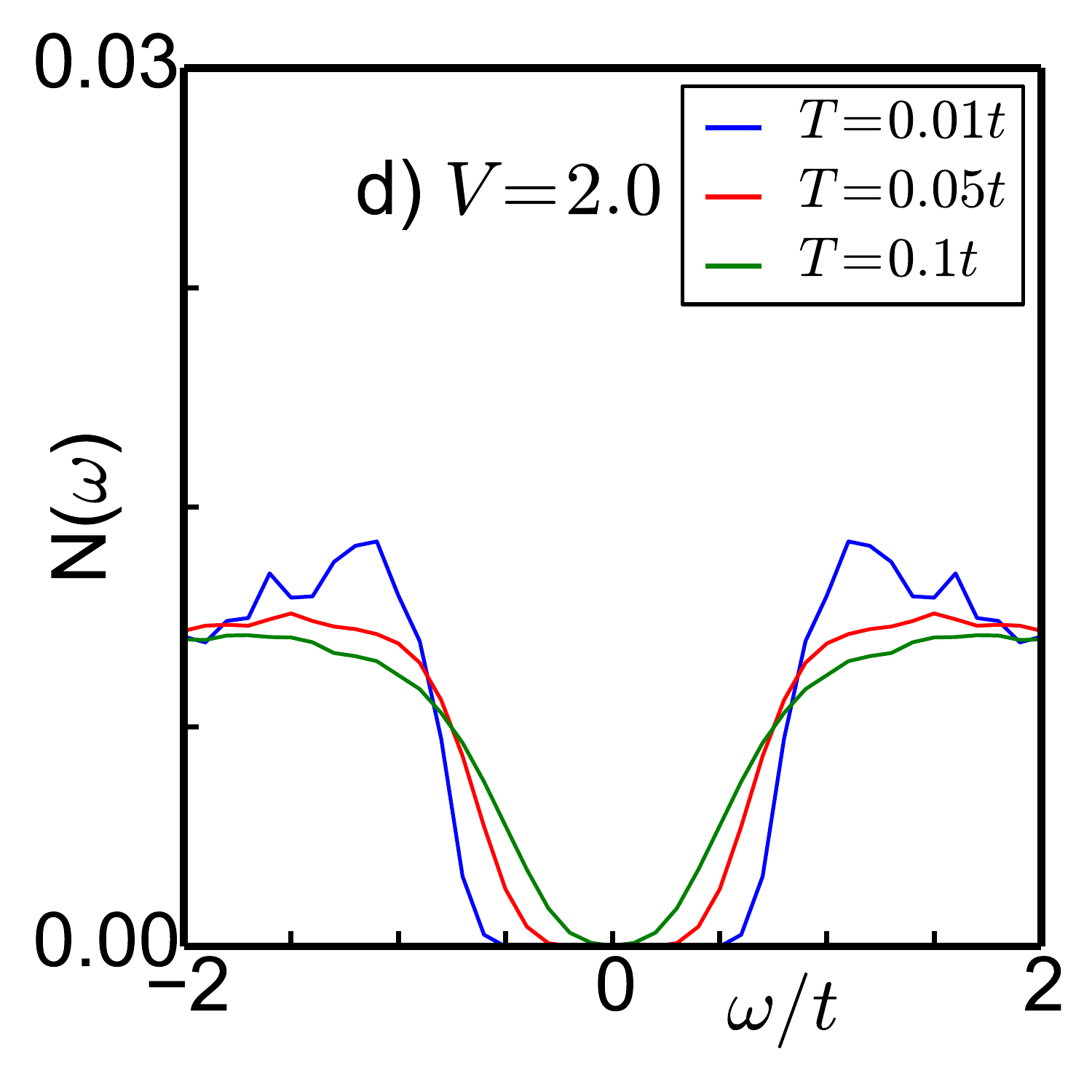}
}
\caption{Disorder and temperature dependence of $\Delta_0.\Delta_i$
at moderate speckle size, $\sigma=2$.  At fixed temperature along the
row the correlation decreases with increase in $V$. The correlation is
strongest at $V=0.5t$ and $T=0.01t$ and decreases with increasing $V$
or $T$.  Bottom row: DOS at $\sigma =2$, showing $V$ and $T$ dependence.
The suppression in coherence peak is seen with increasing temperature
and disorder strength. The gap in the density of states reduces
with increasing temperature. This effect is more pronounced
at small disorder strength.  }
\end{figure*}
%-----------------------------------------------------------------

\subsection{Variation in speckle size}

Fig.6 shows the spatial correlations of the pairing
field at $V =2t$ (where the uncorrelated disorder problem
would be insulating) for four speckle sizes and three temperatures.
To keep track of both amplitude and phase correlations we plot the
scalar $C_i = {\Delta}_0.{\Delta}_i$, where ${\vec R}_0$ is
a reference (corner) site and ${\vec R}_i$ is the site under
consideration, treating $\Delta$ like a two dimensional `vector'.
At the lowest $T$ there is
hardly any pairing correlation at $\sigma=1$, some trace at $\sigma =2$,
and percolative patterns at $\sigma=3-4$.
Naturally the finite $T$ systems at $\sigma =1, 2$ do not have
any SF correlations, the $\sigma =3$ system seems to lose
global correlation at $T < 0.05t$, while the $\sigma =4$
system loses its order somewhere between $T=0.05t$ and $T=0.1t$.
One can draw a $T_c(\sigma)$ plot akin to Fig.5(b)
that shows the onset of SF order for $\sigma > 2$ and
a gradual enhancement of $T_c$ with $\sigma$.
With increase in $\sigma$ the low $T$ gap in the DOS
reduces while the coherence
peak sharpens. Increase in $T$ leads to a quicker closure of
the gap in the large $\sigma$ system.

\subsection{Variation in disorder strength}

Fig.7 shows data that is complementary to Fig.6, now focusing on
results at a fixed speckle size $\sigma=2$. As expected the low $T$
system shows SF order for $V \lesssim 1.5t$ and insulating
character for $V =2t$.
The $V=2t$ system continues to remain insulating at all $T$
while the weaker disorder cases show a fragmentation of the
spatial order, and its loss at a scale $T_c(V)$, with
increasing $T$. The associated $T_c$ are given in Fig.5(a).

The change in DOS with $V$ and $T$ is as anticipated. At low
$V$ the low $T$ DOS has a large gap with sharp edge and reasonable
coherence peaks. With increasing $T$ there is a transfer of weight to
low energy and a smearing of the coherence feature.
With increase in $V$ the low $T$ DOS shows a smaller gap and for
$V \gtrsim 1.5t$ no coherence peaks are visible. However, the
transfer of spectral weight to low frequency, due to increasing $T$,
is weaker in the larger $V$ case compared to weak disorder.

%-----------------------------------------------------------------
\begin{figure*}[t]
\centerline{
\includegraphics[width=12.0cm,height=8.2cm]{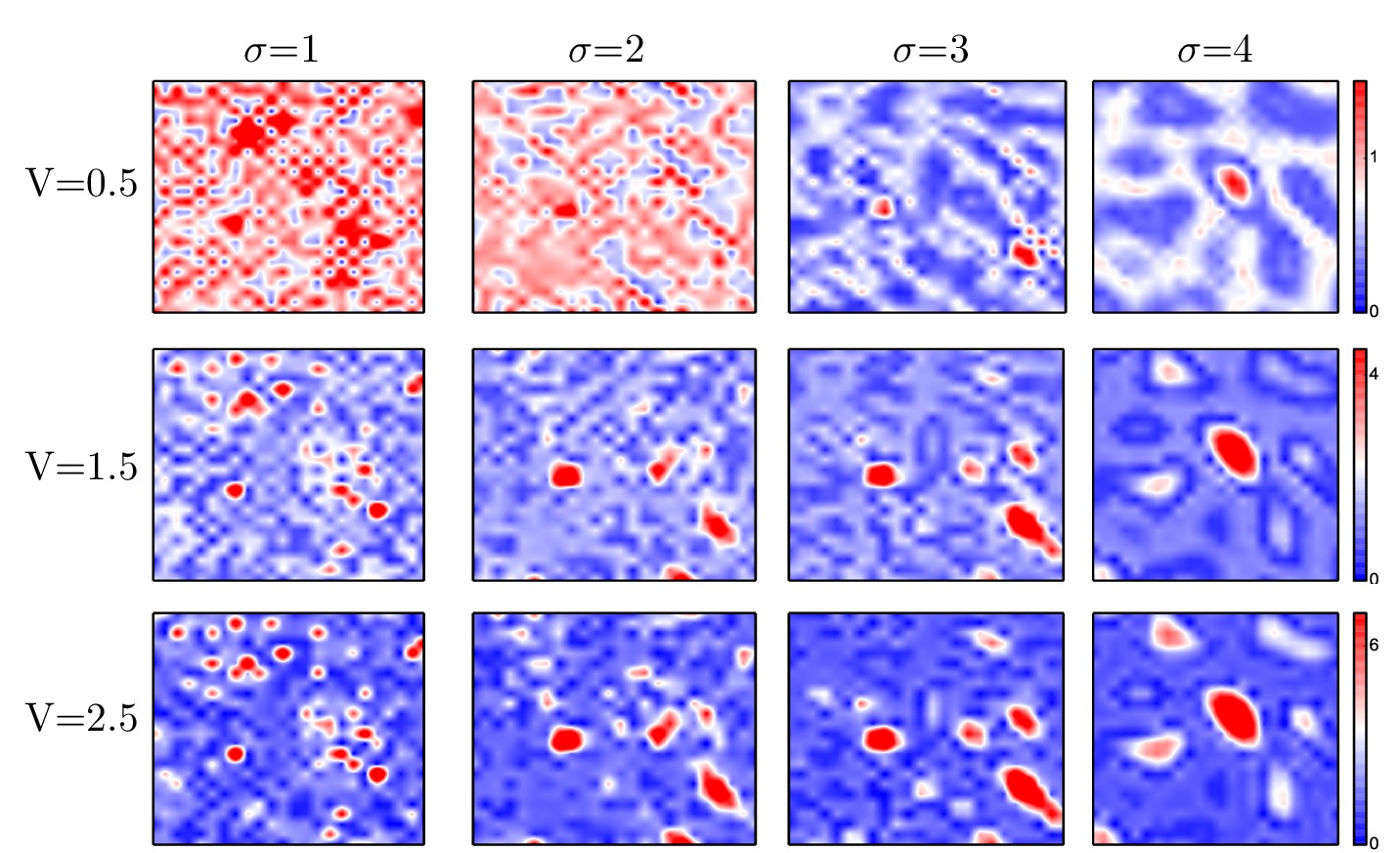} }
\caption{Maps
for the effective potential $V_{eff}$ for varying speckle size
and disorder strength.  The disorder is $V = 0.5t,~1.5t,~2.5t$
from top to bottom, while the speckle size is $\sigma=1,~2,~3,~4$
from left to right.  The bottom left panel - large $V$ and small
$\sigma$ - has the most fluctuating pattern while the top right
panel - small $V$ and large $\sigma$ - has the smoothest profile.
}
\end{figure*}
%-----------------------------------------------------------------

\section{Discussion}

Having seen the results of increasing speckle size on the
ground state and thermal properties of the superfluid, we
want to suggest how these effects arise from 
the renormalised effective potential that emerges 
in this problem.

\subsection{Understanding the ground state}

\subsubsection{Nature of the effective disorder}

From previous studies on uncorrelated disorder\cite{ghosal-prb}
we know that the presence of the Hartee term in HFBdG
Hamiltonian enhances  the effect of disorder. 
At $U \gg t$ and in the presence of disorder the density field
become strongly inhomogeneous due to the Hartree feedback from
the interaction term. As a result, the effective potential to 
which the fermions react is not $V_i$ but
$
V_{eff} = V_i + \phi_i  = V_i + {U \over 2} \langle n_i \rangle. $
In contrast to a weakly interacting system with uncorrelated
disorder the pairing in the present problem would involve
fermions in an effective potential that is (i)~strongly
renormalised due to the large $U$, and (ii)~spatially
correlated due to the fermionic feedback and finite
$\sigma$. The combination of $V$ and $U$ enhances
localisation, while increasing $\sigma$ at fixed $(V, U)$
weakens localisation.
These effects in turn impact on the phase stiffness which
dictates the $T_c$ scales of the superfluid \cite{tarat-arx}.

With this in mind, Fig.8 shows maps of the effective potential
$V_{eff}$ for varying $V$ and $\sigma$. The $\phi$ that enters
$V_{eff}$ is obtained via the full HFBdG minimisation.
As expected the $V_{eff}$ at small $\sigma$ is rapidly fluctuating
while at large $\sigma$ the variation is much smoother. The main
impact of $\phi_i$ is to increase the width of the effective disorder
leaving the spatial correlation more or less as in $V_i$. This
is borne out by comparing  Fig.8 with the top row
in Fig.2.

We show the distribution of $V_{eff}$ at three values of $V$
in  Fig.9(a)-(c), comparing results at $\sigma =1$ and $\sigma=4$.
The distributions show marginally greater weight at 
large $V_{eff}$ for the larger $\sigma$ case.The inset of Fig.9(c) compares the
standard deviation $\delta V_{eff}=\sqrt{\braket
{V_{eff}(\vec{r})^2}-\braket{V_{eff}(\vec{r})}^2}$
for two values of $\sigma$. The $\delta V_{eff}$  at
$\sigma=4$ is only slightly larger than that at $\sigma=1$.
We conclude that the local distribution of $V_{eff}$ is mainly
independent of speckle size.

%-----------------------------------------------------------------
\begin{figure}[b]
\centerline{
~
\includegraphics[width=4.2cm,height=3.3cm]{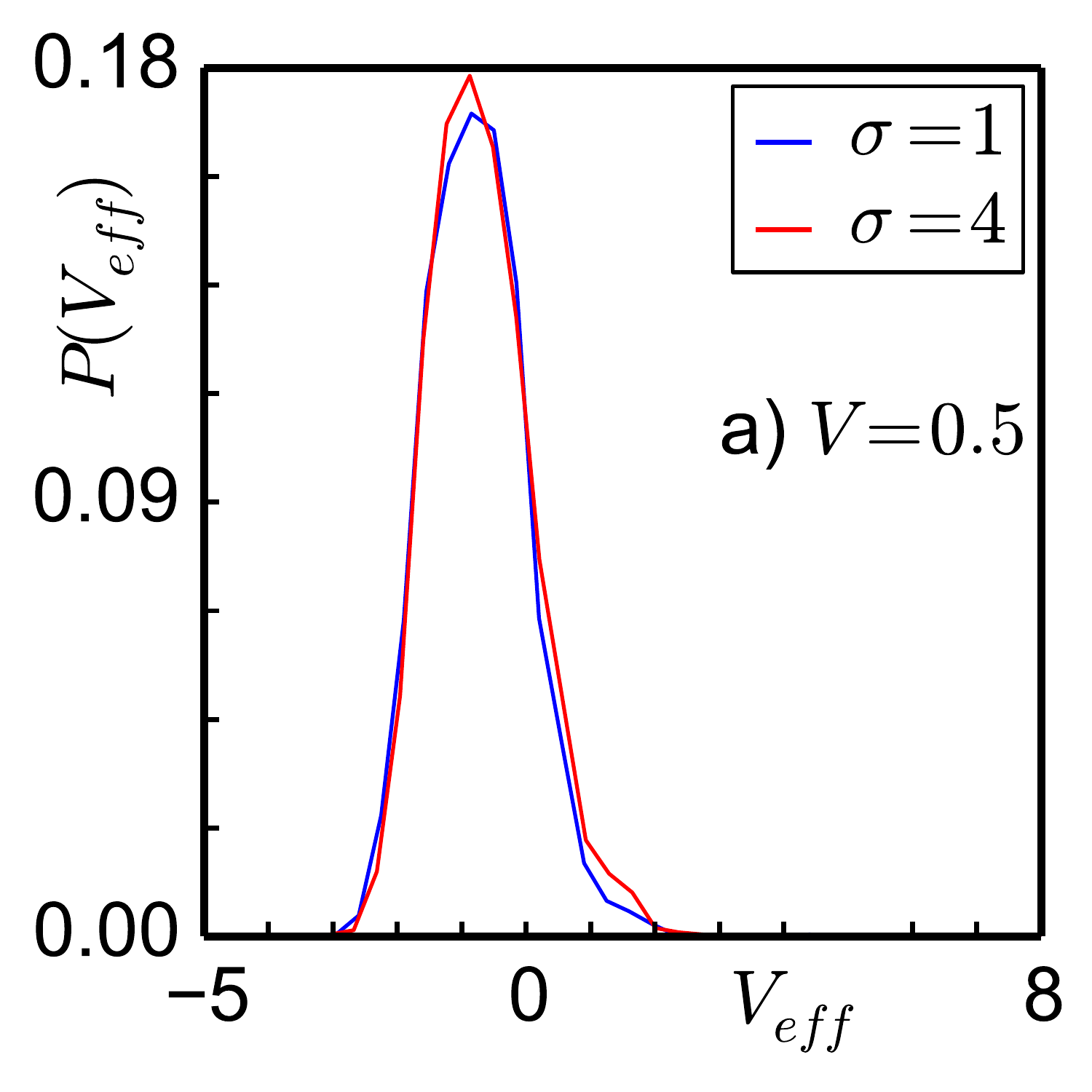}
~\includegraphics[width=4.2cm,height=3.3cm]{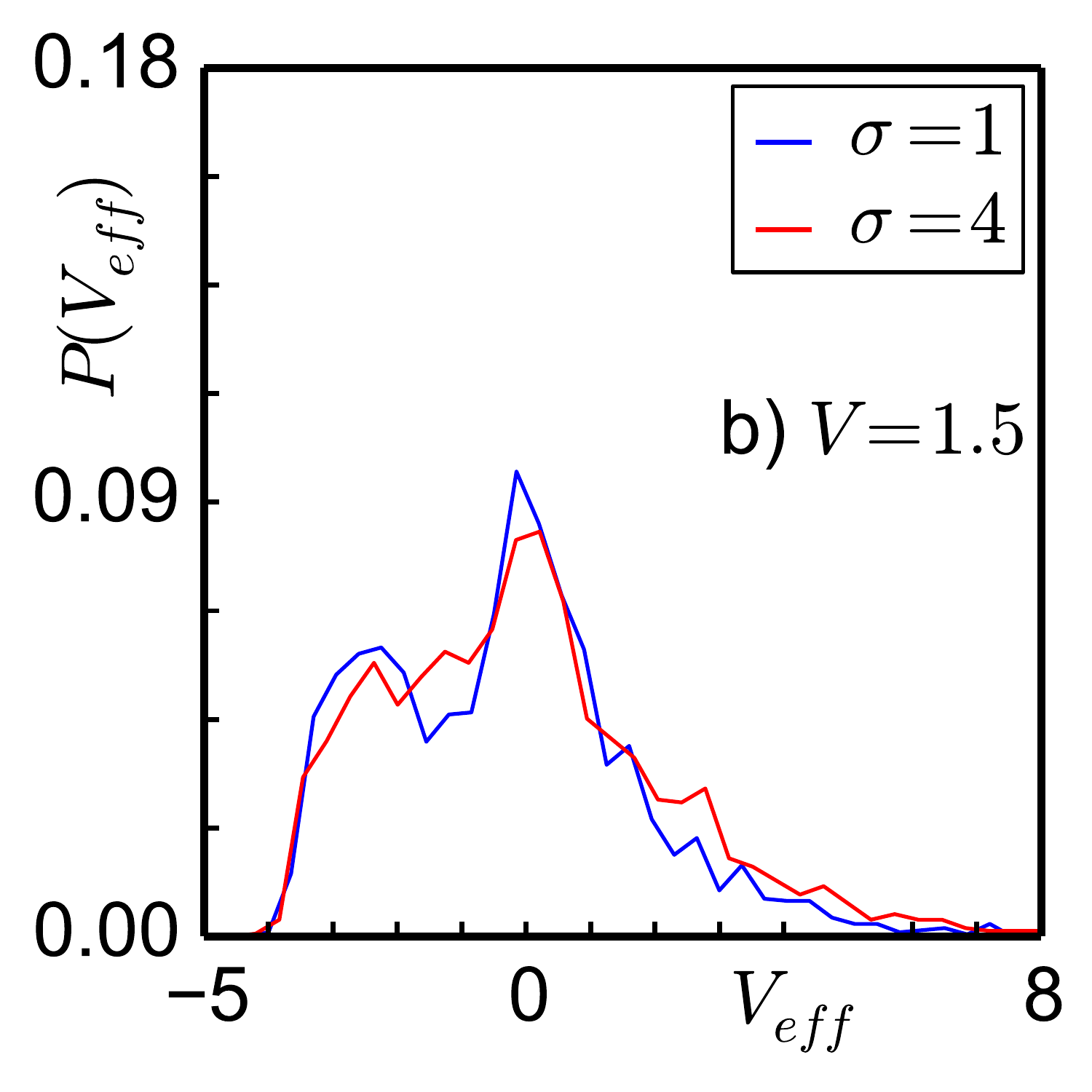}
}
\centerline{
\includegraphics[width=4.2cm,height=3.5cm]{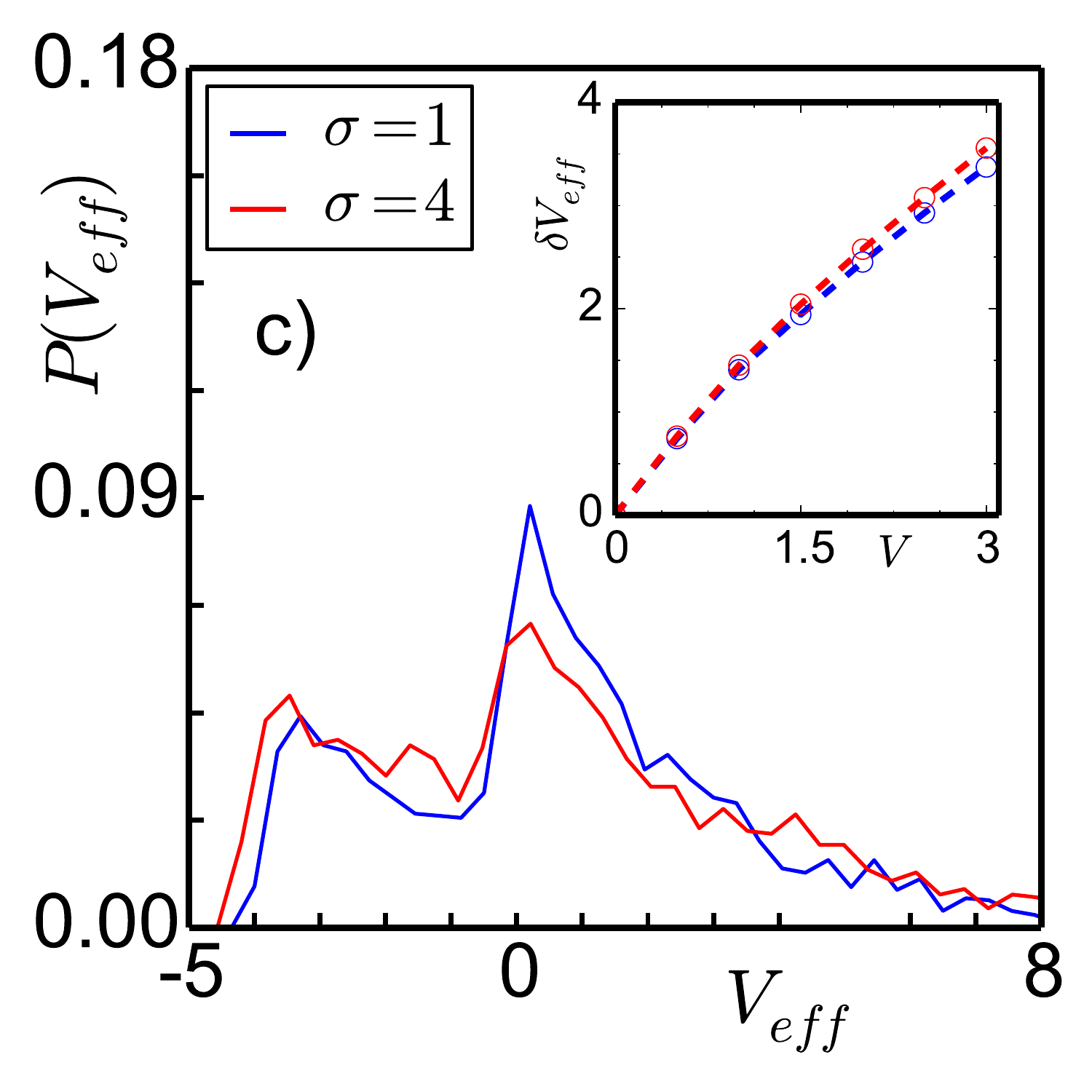}
~\includegraphics[width=4.1cm,height=3.5cm]{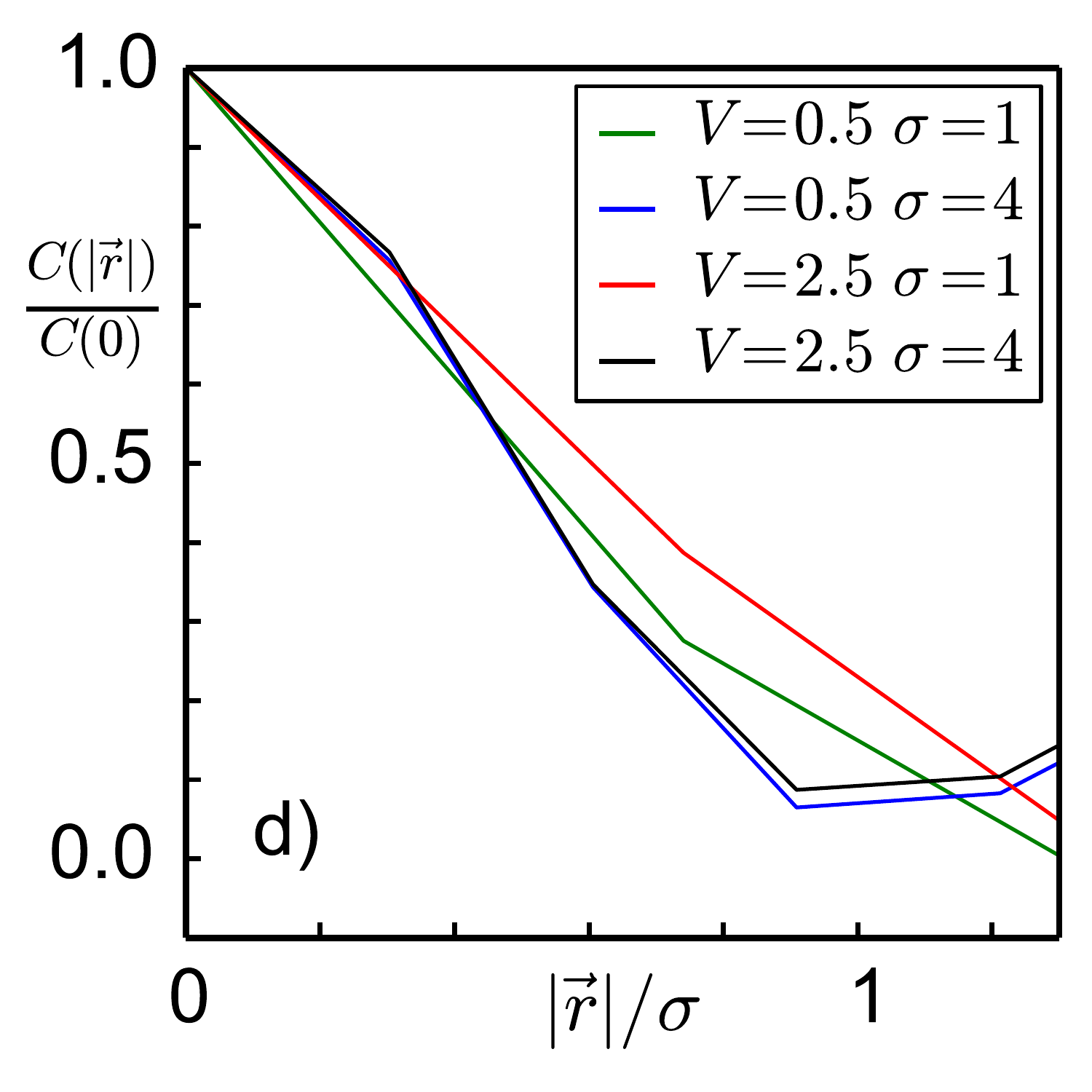} }
\caption{(a)-(c) shows $P(V_{eff})$ at $V=0.5t,1.5t,2.5$,
respectively, comparing $\sigma=1$ with $\sigma =4$.
Inset of Panel (c) shows the variance $\delta V_{eff}$ with respect to
$V$ at $\sigma=1,4$. At a given $V$ the variance is slightly
larger at $\sigma=4$ compared to $\sigma=1$.  (d)~The
normalised $C(\vec{r})$ as a function of disorder for two speckle
sizes. The behaviour suggests that the spatial correlation in
$V_{eff}$ is dictated by just $\sigma$ and is unaffected $V$.
}
\end{figure}
%-----------------------------------------------------------------
%-----------------------------------------------------------------
\begin{figure}[t]
\centerline{
\includegraphics[width=2.8cm,height=3.5cm]{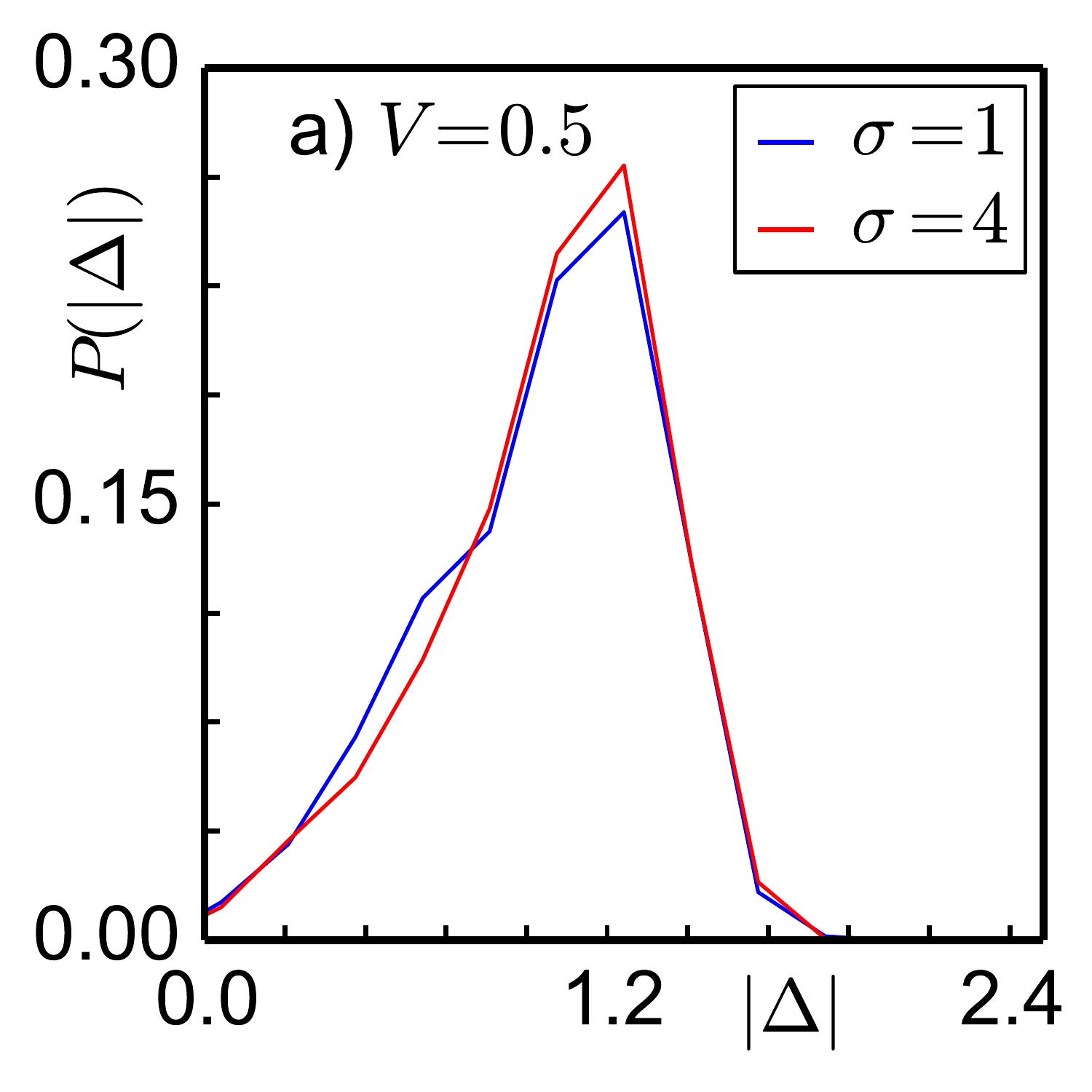}
\includegraphics[width=2.8cm,height=3.5cm]{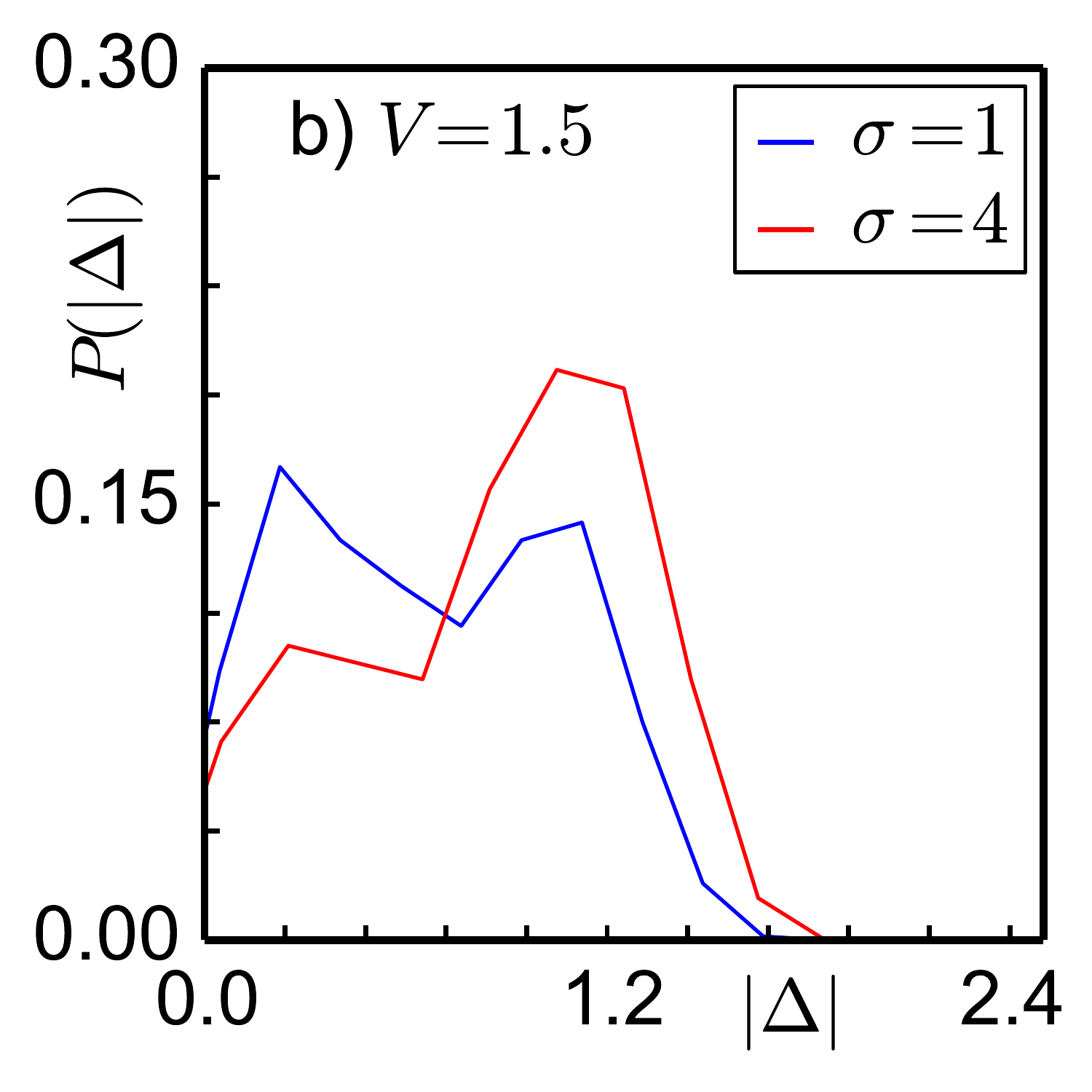}
\includegraphics[width=2.8cm,height=3.5cm]{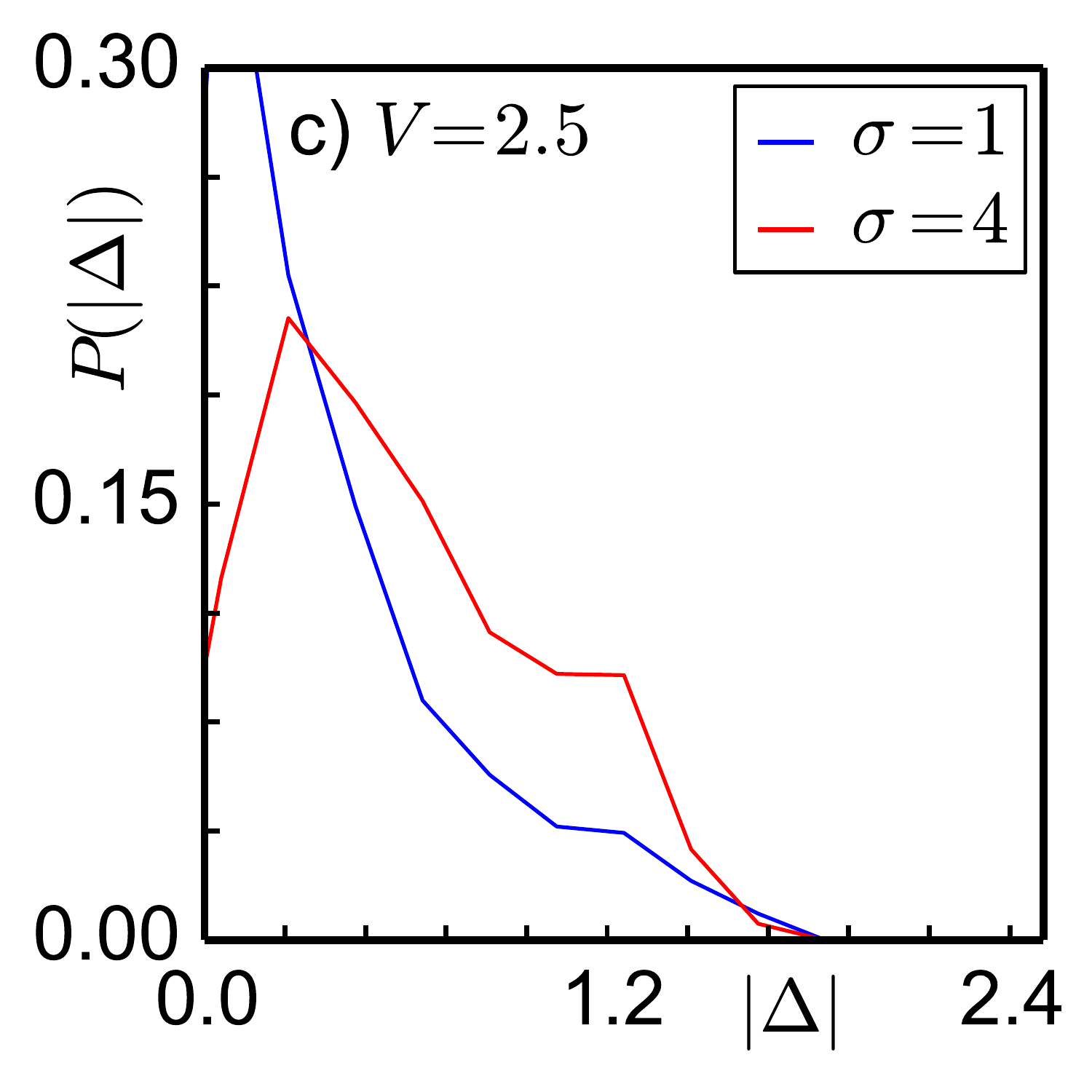}
}
\caption{
The distribution of $\vert \Delta_i \vert$ in the ground state
comparing the $\sigma$ dependence at $V=0.5t,1.5t,2.5t$. At
larger $V$ the effect of $\sigma$ is to create a
distribution with large weight at low amplitude.
}
\end{figure}
%-----------------------------------------------------------------

To characterise the spatial correlations in $V_{eff}$ 
panels 9(d)  show the plot of
$$
C(|\vec{r}|)
=\braket{V_{eff}(\vec{x}+\vec{r})
V_{eff}(\vec{x})}-\braket{V_{eff}(\vec{x})}^2
$$
for speckle sizes 1 and 4. These indicate that the spatial
correlations depend on $\sigma$ but are essentially $V$
independent.
Overall, Fig.9 suggests that the strength of $V_{eff}$ is 
dictated by $V$ and $U$ while the correlations in $V_{eff}$ 
are dictated by $\sigma$ only.

%-----------------------------------------------------------------
\begin{figure}[b]
\centerline{
~
\includegraphics[width=4.2cm,height=3.5cm]{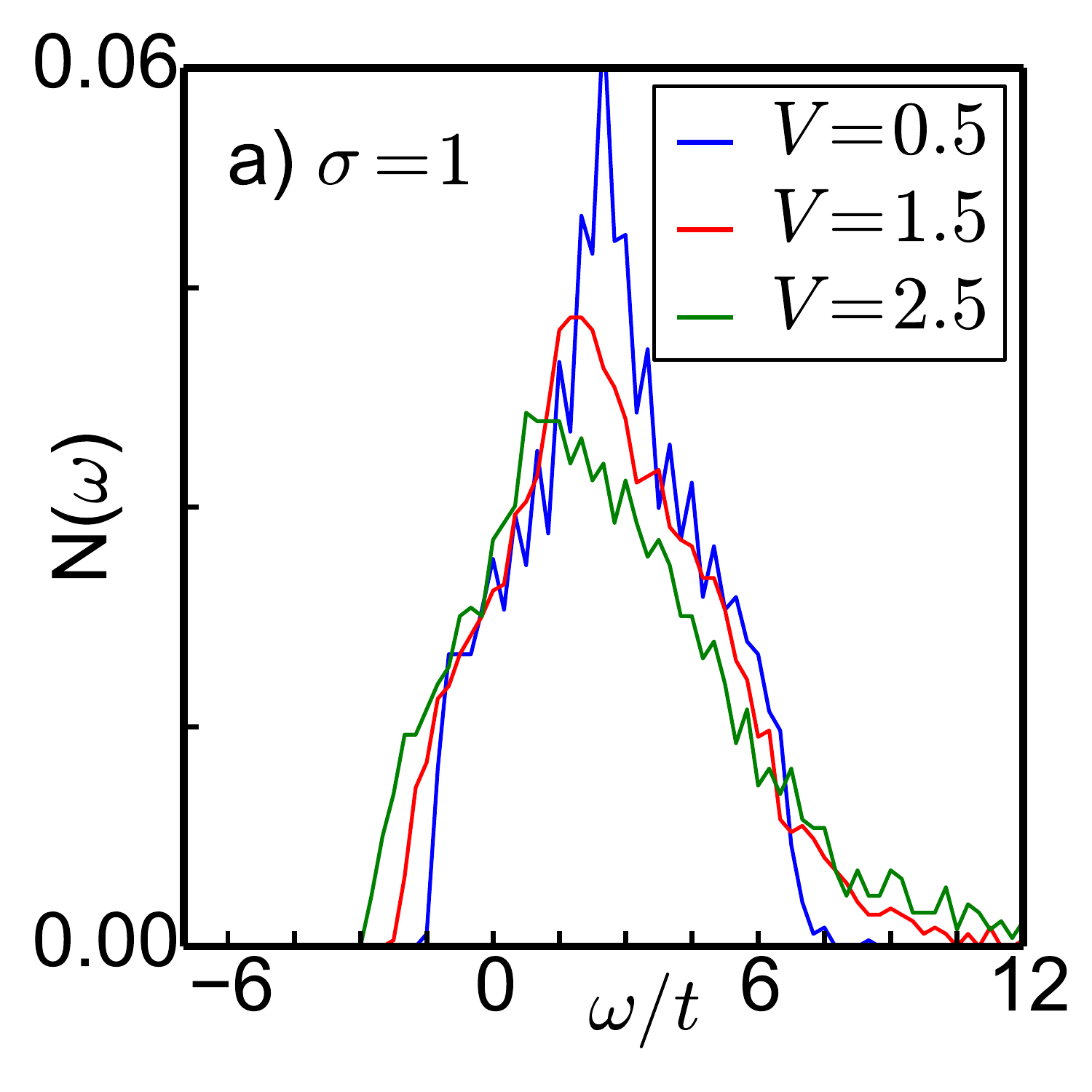}
\includegraphics[width=4.2cm,height=3.5cm]{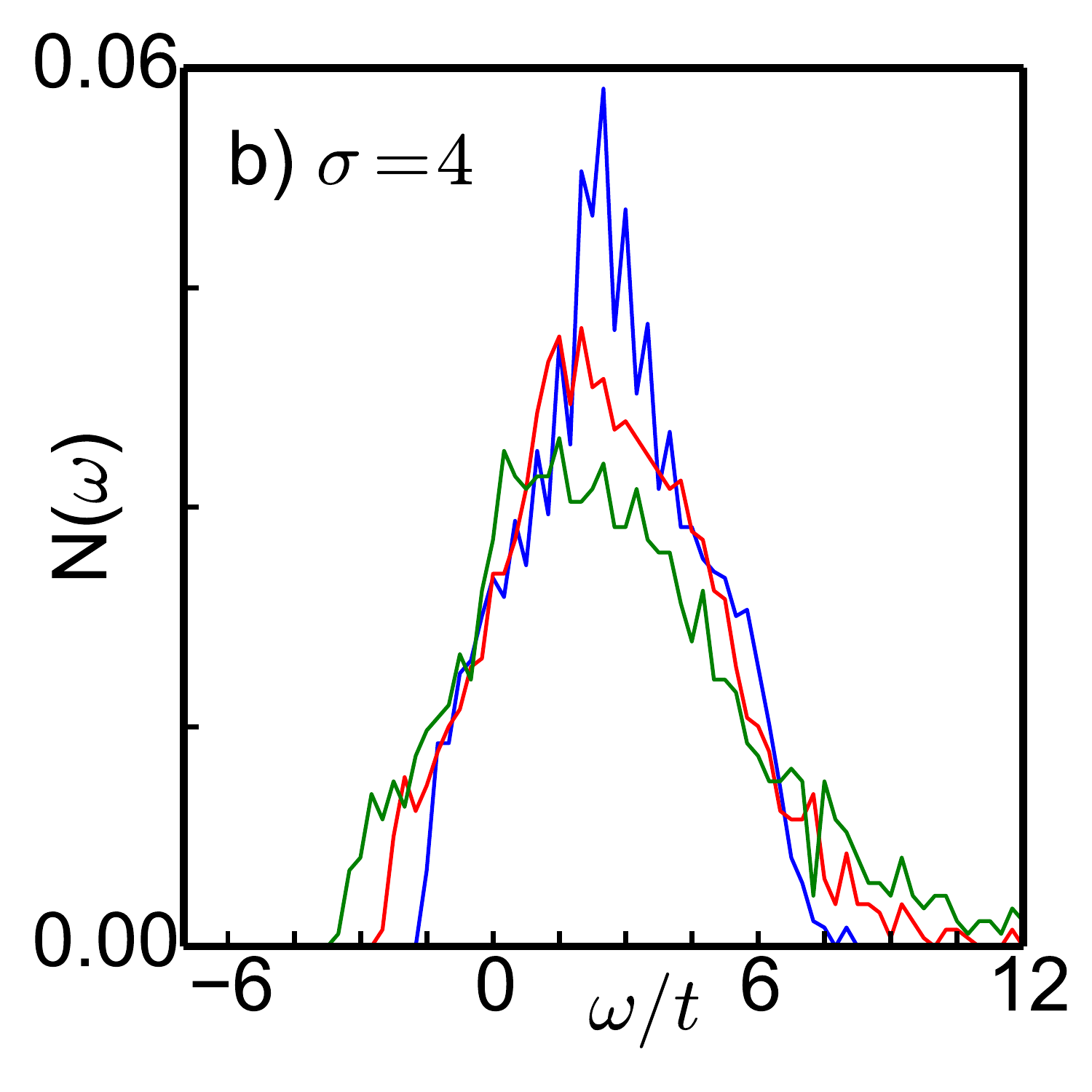}
}
\centerline{
\includegraphics[width=4.2cm,height=3.5cm]{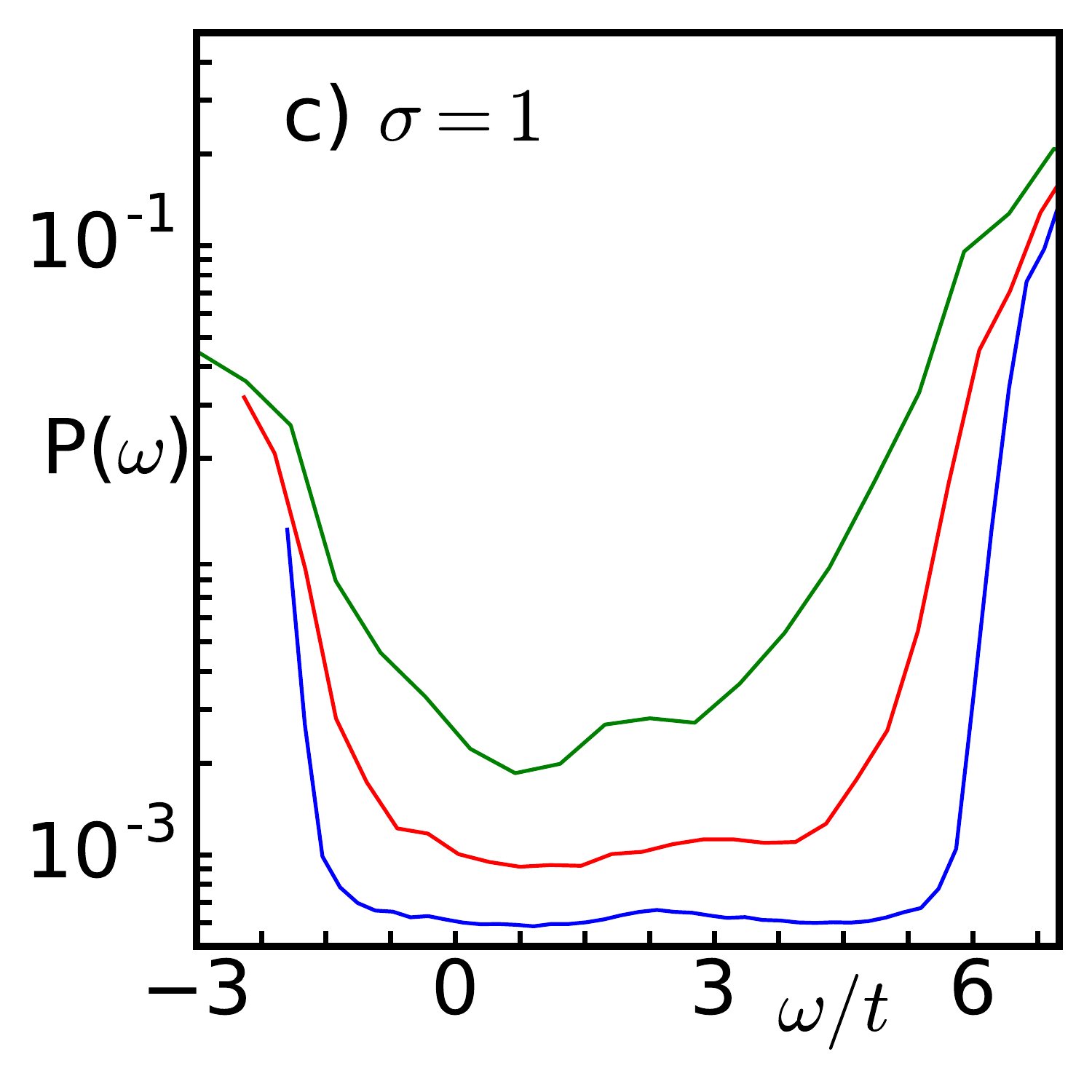}
\includegraphics[width=4.2cm,height=3.5cm]{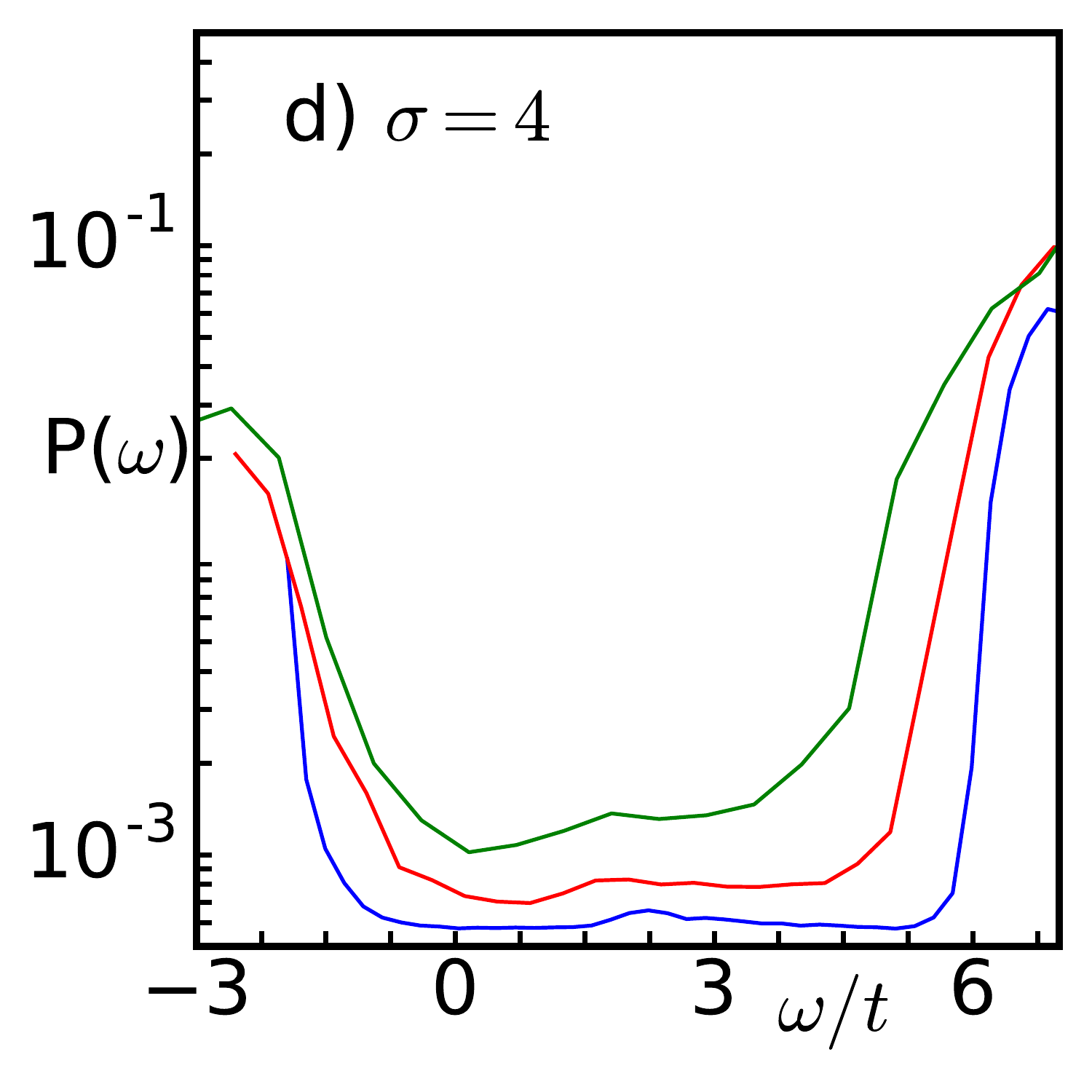} }
\caption{DOS and IPR in the background of the bare disorder, $V_i$.
The DOS naturally broadens with $V$ but is not very sensitive to
$\sigma$. The IPR however is sensitive to $\sigma$ and shows weaker
localisation (smaller  IPR) at larger $\sigma$.
}
\end{figure}
%-----------------------------------------------------------------
%-----------------------------------------------------------------
\begin{figure}[t]
\centerline{
\includegraphics[width=4.2cm,height=3.5cm]{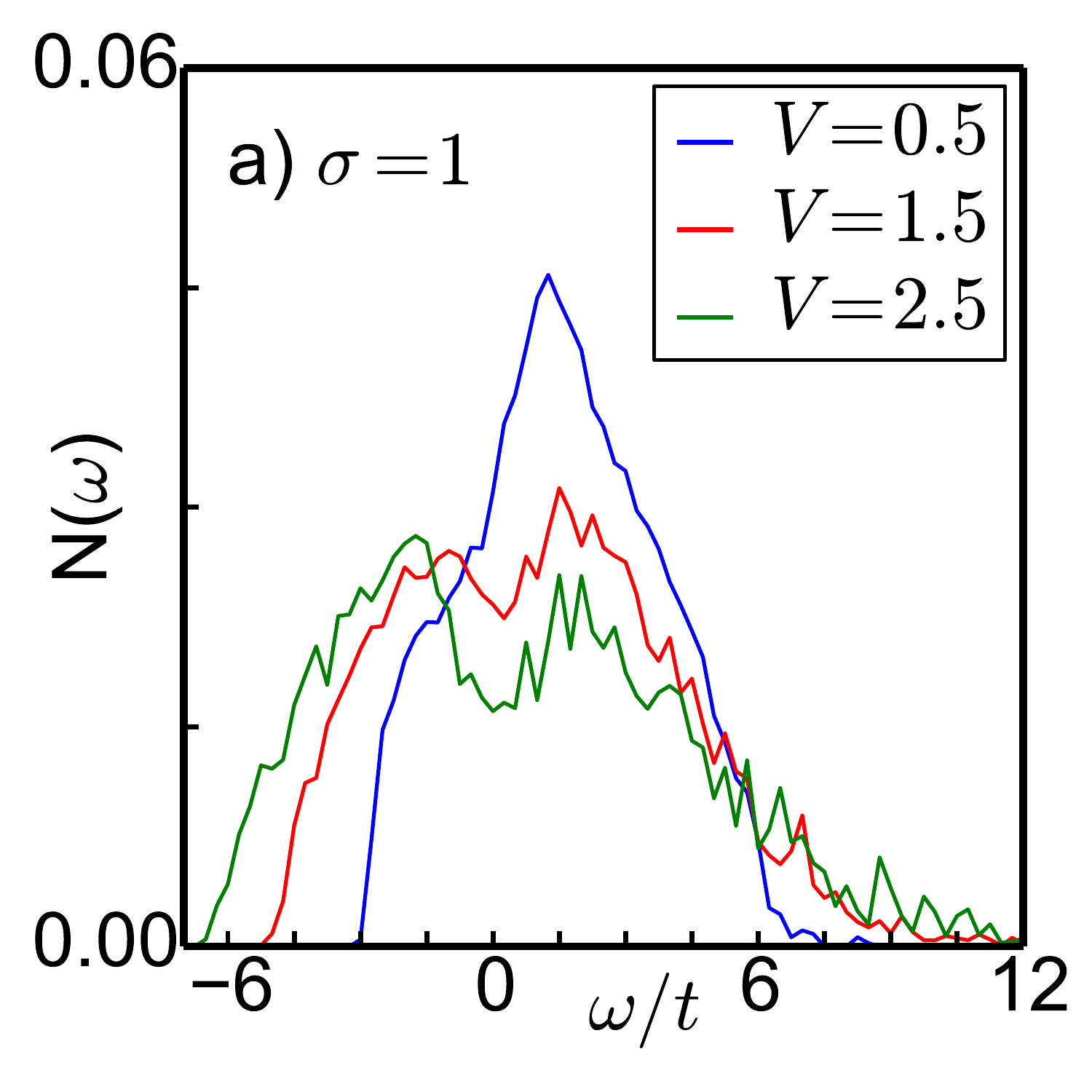}
\includegraphics[width=4.2cm,height=3.5cm]{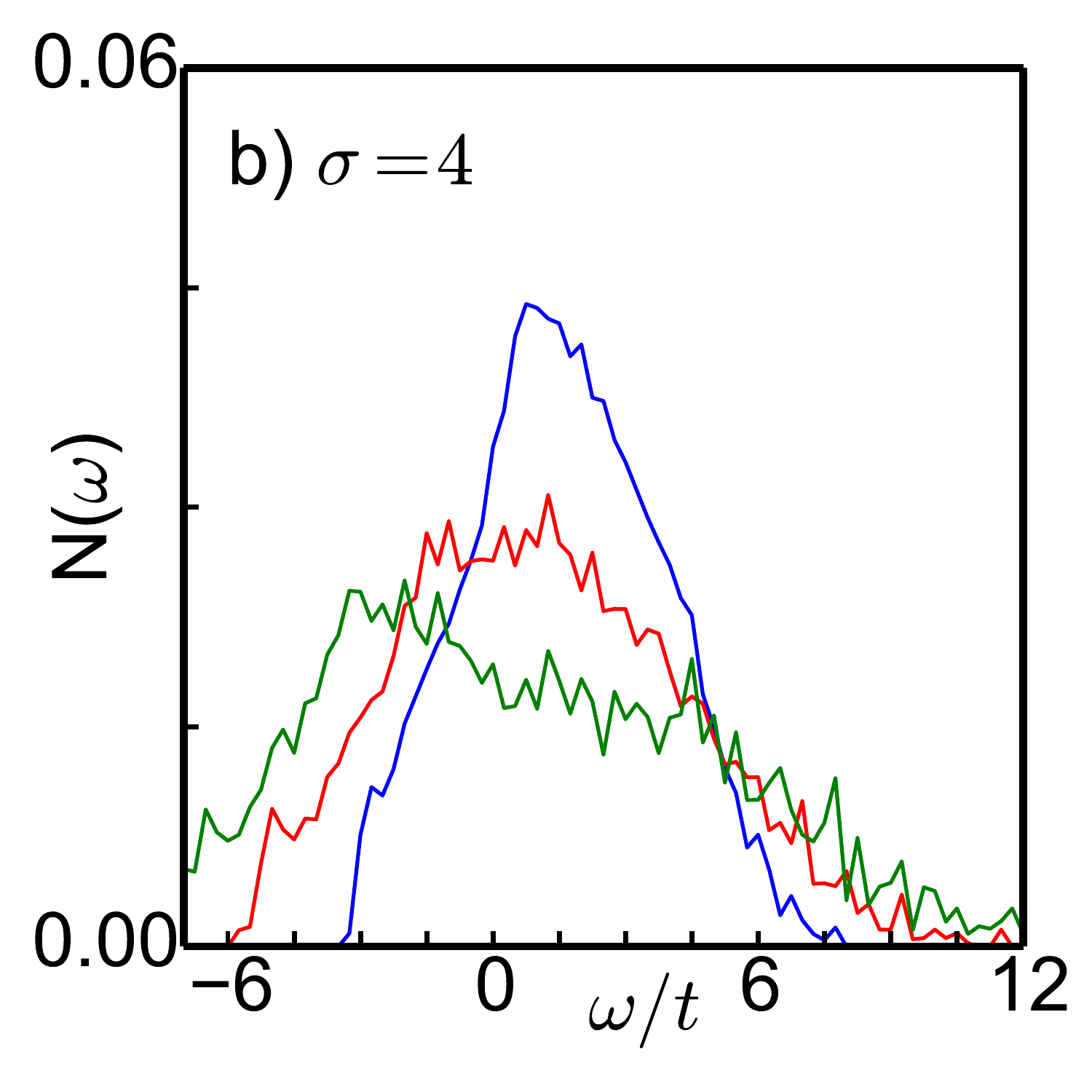}
}
\centerline{
\includegraphics[width=4.2cm,height=3.5cm]{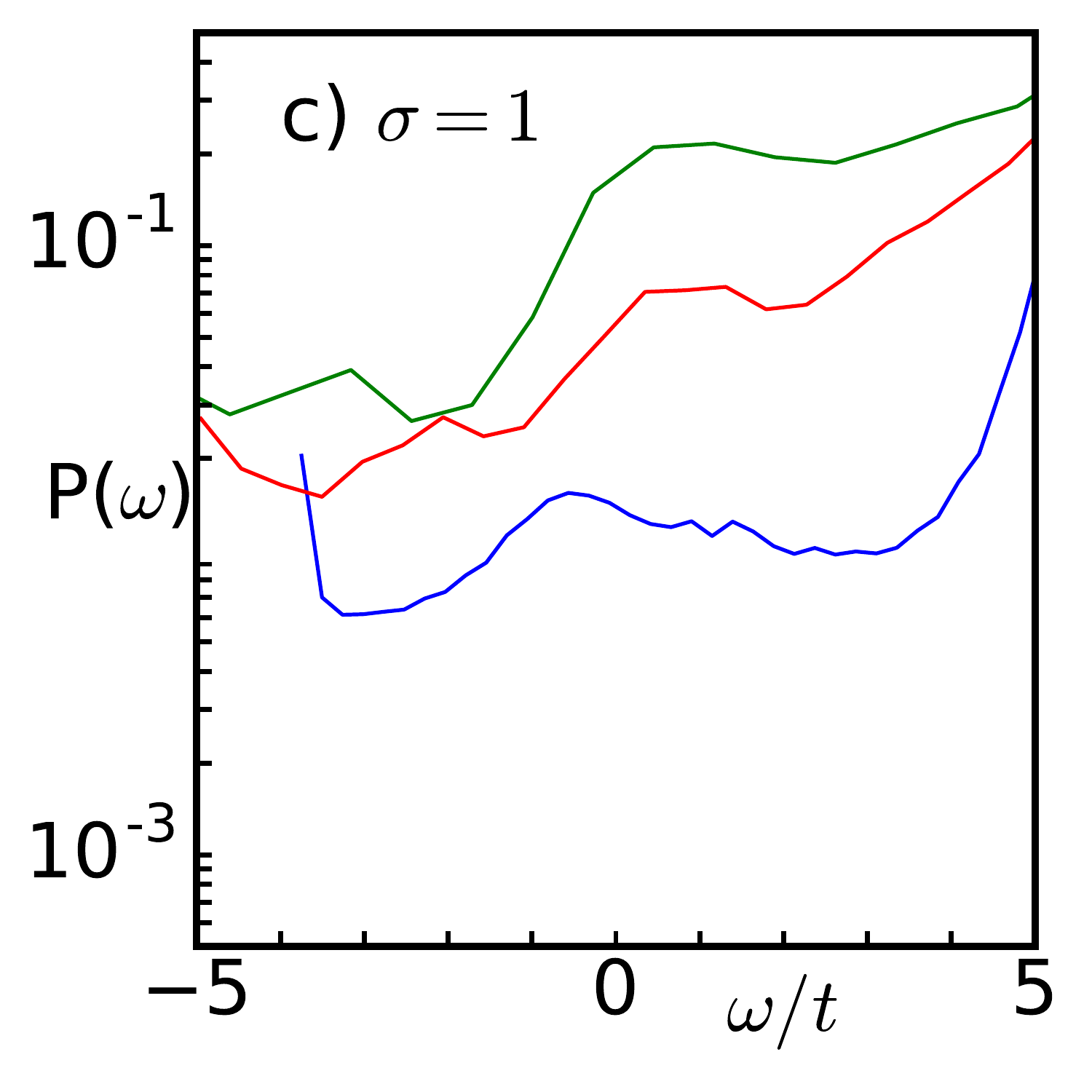}
\includegraphics[width=4.2cm,height=3.5cm]{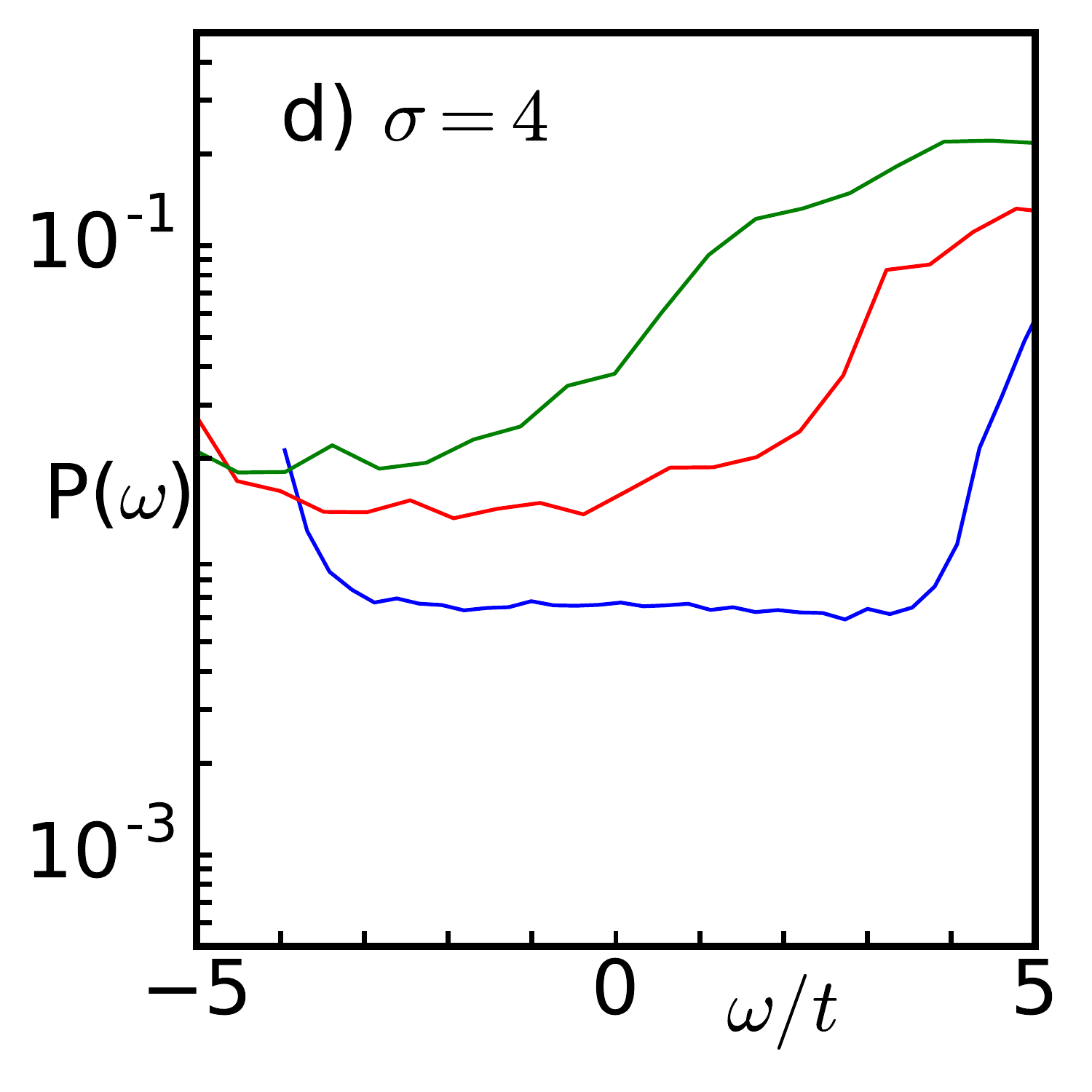} }
\caption{DOS and IPR in the presence of the effective disorder
$V_i^{eff} = V_i + \phi_i$.  We have ignored the pairing effects
in this calculation.  The $V_i^{eff}$ problem has larger bandwidth,
due to the larger effective disorder, and greater localisation
compared to the bare disorder. The IPR is much larger than in the
$V_i$ problem, and much larger at $\sigma=1$ compared to
$\sigma =4$.  }
\end{figure}
%-----------------------------------------------------------------

\subsubsection{Order parameter in the ground state}

Fig.10 shows the distribution of $\vert \Delta_i \vert$
in the ground state for three values of $V$ and two
speckle size. While $\sigma$ has little effect in
the distribution at $V=0.5t$, there is a distinct $\sigma$
dependence at larger disorder - for a given $V$ the distribution
at larger $\sigma$ has much greater weight at low amplitude.
This correlates with the behaviour of the spatial patterns
and the DOS that we have seen earlier.

\subsubsection{Localisation effects}

While the $V_{eff}$ and the resulting $\Delta_i$ 
control features like the DOS, to understand 
intersite coupling between the $\Delta_i$ we
need to understand the spatial extent of the
wavefunctions in the $V_{eff}$ background.

First the case of bare disorder, to set a reference.
Fig.11 shows the DOS (upper row) and the IPR (lower row) 
in the presence of only $V_i$. 
The model is solved with $U$ set to zero.
We show results for $V=0.5t,1.5t,2.5t$ and $\sigma=1,4$.
Since $P(v) \sim e^{-v/V}$ has a finite positive mean the
DOS and IPR plots are asymmetric about $\omega=0$. The greater
width of the disorder distribution at larger $V$ leads
to a correspondingly broader DOS. Comparing panels
(a) and (b) the speckle
size does not make a significant difference to the DOS.
The IPR shows a more significant $\sigma$ dependence,
particularly at large $V$. At $V=0.5t$ the IPR at the
band center is $\ll 10^{-3}$, suggesting a 
localisation length $> \sqrt{10^3}$, larger than
our system size $(24 \times 24)$. 
At $V=2.5t$, however, there is a visible
difference between the band center IPR at $\sigma=1$ and
$\sigma =4$. Nevertheless the numbers for the IPR are
still $\sim 10^{-3}$, indicating a large localisation 
length.

Fig.12 shows results on the DOS and IPR based on a $V_{eff}$
extracted from the solution of the HFBdG equation. The
effective model that is solved to obtain the results in
the figure is
$$
H = -t\sum_{<ij>} c^\dagger_ic_j +\sum_i V^{eff}_i n_i
$$
As in Fig.11 the disorder values are $0.5t,~1.5t,~2.5t$.
The fermions are subject to a larger
effective disorder than in Fig.11. As a result the
weight spreads over a
larger frequency window. Here again the the effect of disorder
is somewhat weaker in (b) compared to (a).

The most interesting feature is the contrast between the IPR
in Fig.11 with that in Fig.12.
Note the following: 
(i)~the IPR in the $V_{eff}$
problem, near $\omega =0$ or near the band edges, is at least an
order of magnitude larger than the corresponding value in the
bare disorder problem, and (ii)~between Fig.12(c) and Fig.12(d) 
the larger
$\sigma$ case shows a clearly smaller IPR. For
example around $\omega =0$ the IPR at $V=2.5t$ and $\sigma=1$
is $\sim 2 \times 10^{-1}$, while at $\sigma=4$ the 
corresponding IPR is $\sim 3 \times 10^{-2}$. The
associated `localisation length' would be $\sim 2$ lattice
spacings at $\sigma =1$ and $\sim 5$ lattice spacings at
$\sigma = 4$. These are well below our system size, and 
significantly different from one another.
This feature of the single particle eigenfunctions enters
the fermionic Green's function
$G_{ij}(i\omega_n)$ and through that the phase stiffness and
$T_c$ scales. We discuss this next.

\subsection{Estimating the phase coupling and $T_c$ scales}

To understand the intersite coupling between the pairing
fields we derive an effective XY model and benchmark it with 
respect to full MC results. A study of the couplings
$J_{ij}$ of this model with disorder and speckle size 
provides some insight on the phase transitions we 
observe in the parent problem.
Postponing a detailed justification to the Appendix, the
approximate model we use is of the form:
\begin{eqnarray}
H_{XY} &=& -\sum_{ij} J_{ij} cos(\theta_i-\theta_j) \cr
J_{ij} ~& = & ~J_{ij}^0\Delta_{i0}\Delta_{j0} \cr
~~\cr
J_{ij}^0 ~& = & {1\over\beta}
\sum_n [F_{ij}(i\omega_n)F_{ji}(i\omega_n) 
+ G_{ij\uparrow}(i\omega_n)G_{ij\downarrow}(-i\omega_n)] 
\nonumber
\end{eqnarray}

%-----------------------------------------------------------------
\begin{figure}[b]
\centerline{
\includegraphics[width=4.2cm,height=4.0cm]{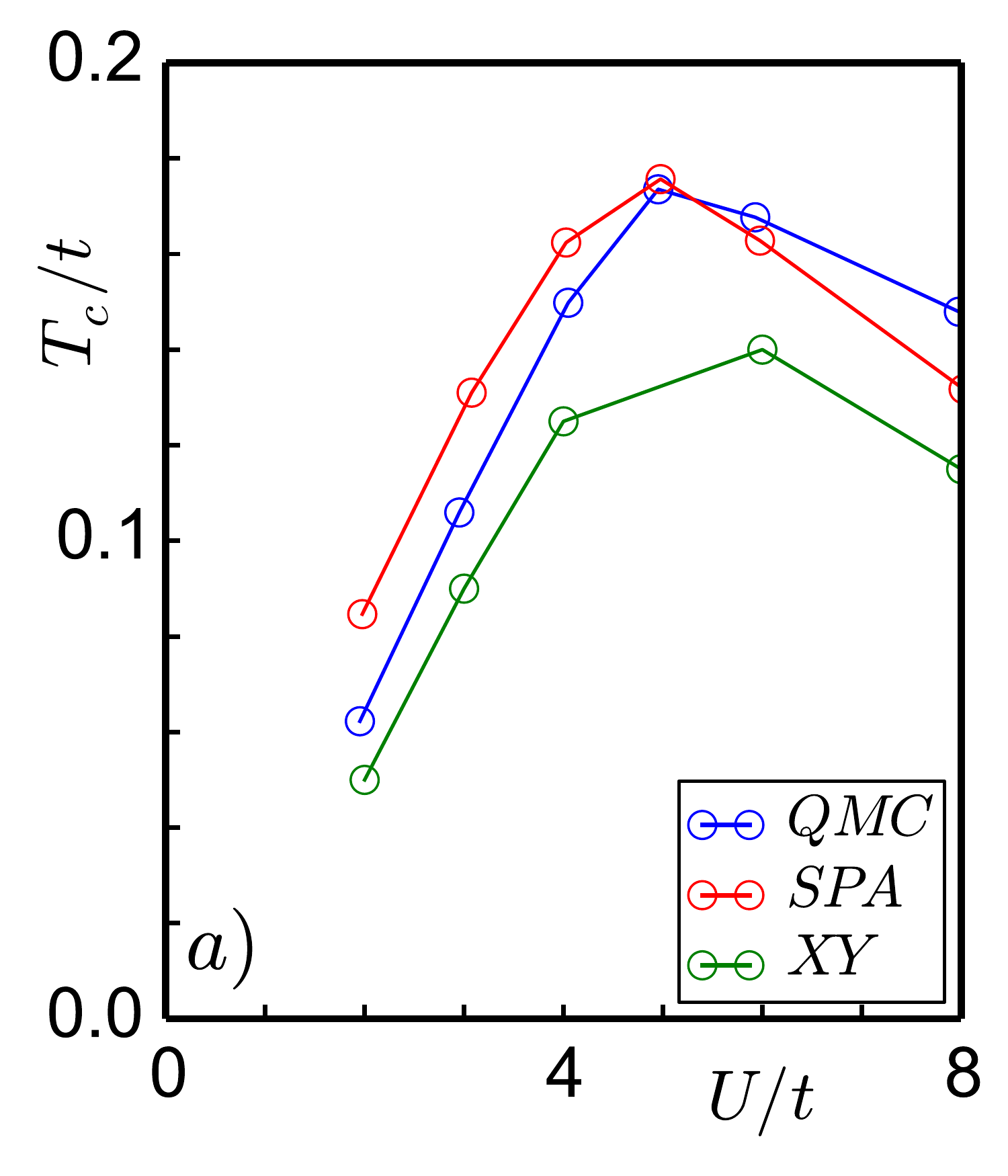}
\includegraphics[width=4.2cm,height=4.0cm]{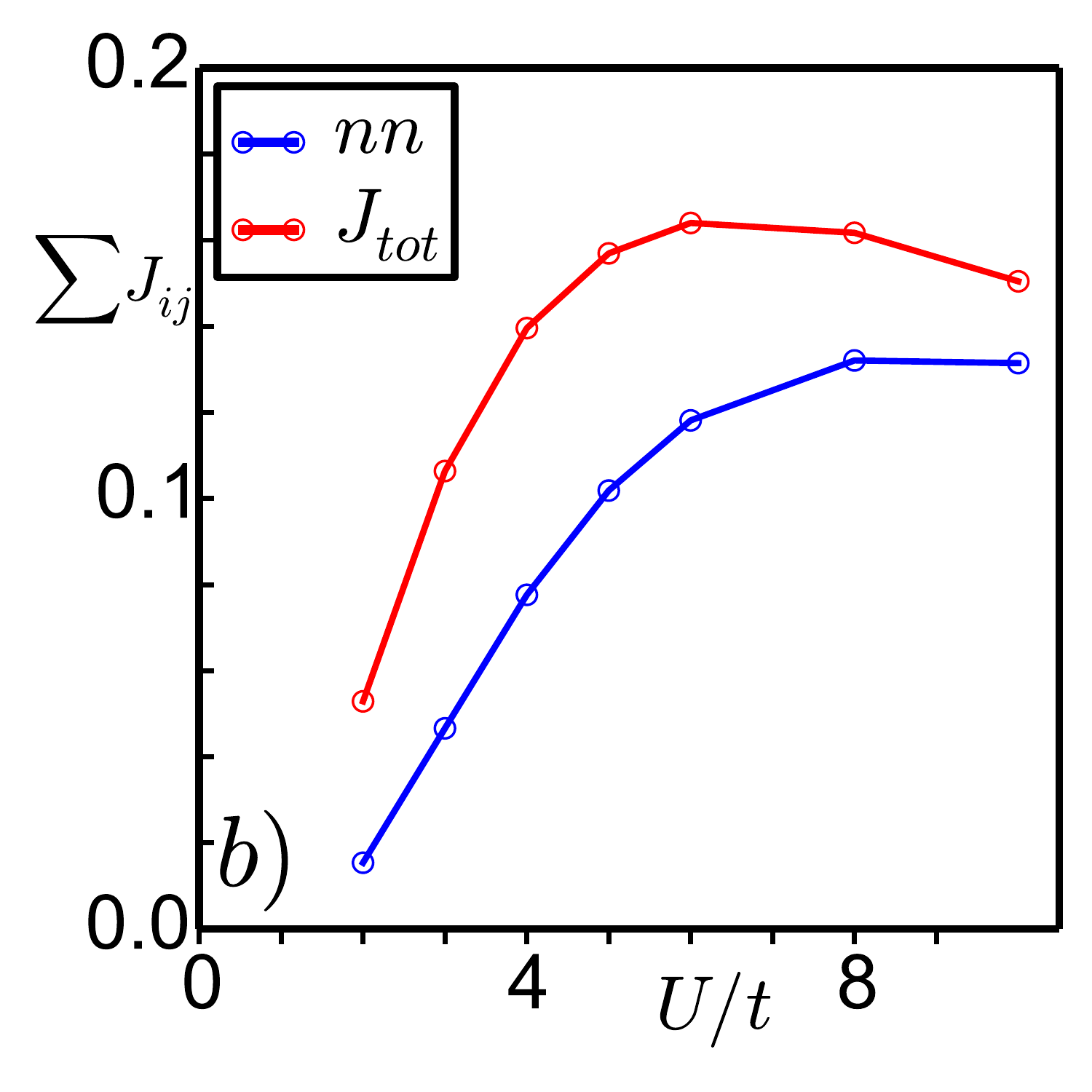}
}
\caption{
(a) The comparison of $T_c$ scales in the clean limit, between
full QMC, our MC result (SPA) and the XY model in the text. 
We operate near the peak $T_c$, the BCS-BEC crossover regime.
The SPA well approximates the QMC answer. 
The XY model also succeeds in capturing the non-monotonic
dependence of $T_c$ on $U/t$. (b) Sum of nearest neighbour XY 
couplings
versus sum of all couplings including nearest neighbour.  For small
to moderate $U$ the model has significant long range couplings and it
is only at very large $U$ that it can be truncated to nearest neighbour.
}
\end{figure}
%-----------------------------------------------------------------
%-----------------------------------------------------------------
\begin{figure}[t]
\centerline{
\includegraphics[width=4.2cm,height=4.0cm]{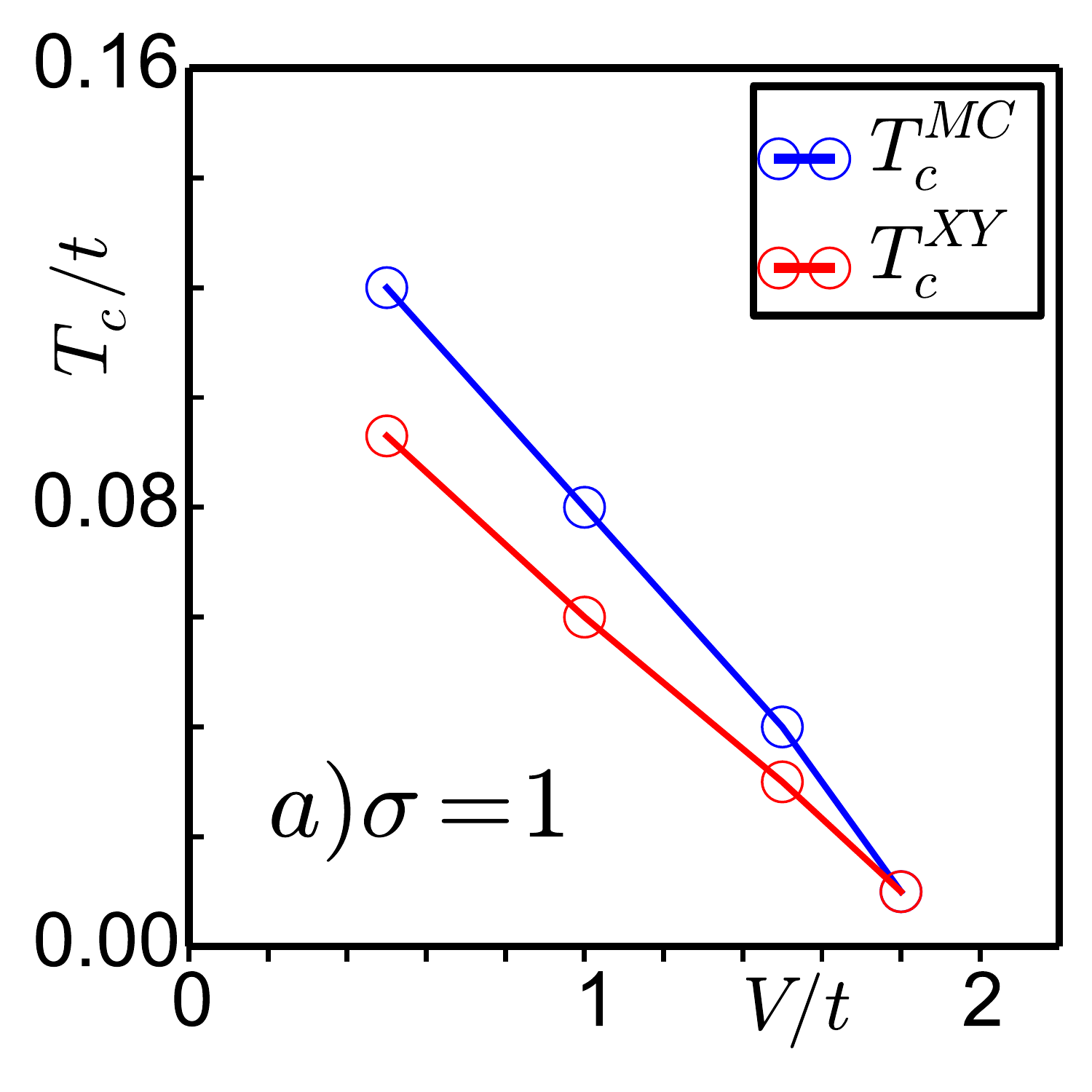}
\includegraphics[width=4.2cm,height=4.0cm]{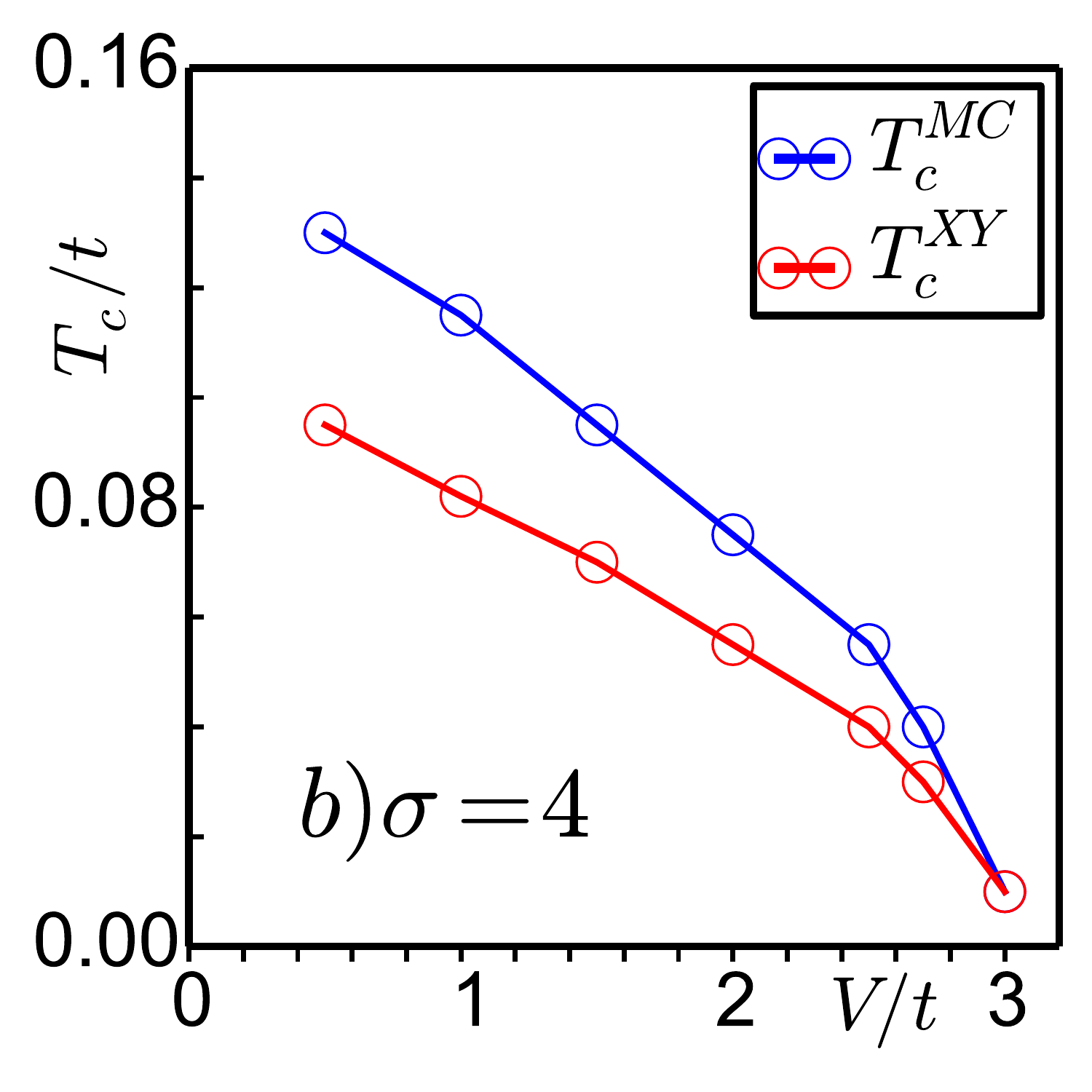}
}
\caption{
Comaparison of $T_c$ scales obtained from exact MC with the XY         
model for two speckle sizes.  Fig a) shows the comparison for speckle
size 1 and Fig b) shows the comparison for speckle size 4.  The
difference in the results of the two calculation increases with
decrease of disorder.  
 }
\end{figure}
%-----------------------------------------------------------------

The $\Delta_{i0}$, etc, are the pairing field amplitude
in the $T=0$ HFBdG state.  
$G_{ij}$ is the `normal' Green's function and 
$F_{ij}$ is the `anomalous' Green's function computed
on the HFBdG state.  Via these Green's
functions $J_{ij}$ 
contains information about {\it excitations} on the 
HFBdG ground state. 
Note that $J_{ij}$ is not limited to nearest neighbours.

First a benchmark in the `clean' problem.  Fig.13(a) 
compares the $T_c$ scales obtained from the full
MC with results from $H_{XY}$ as $U/t$ is varied across
the BCS to BEC crossover.
Given that no explicit finite temperature corrections have
been included in the parameters of $H_{XY}$, the match is
reasonable - and gets better at large $U$.
Fig.13(b) focuses on the couplings that contribute to the
$T_c$. At weak to intermediate coupling, in this clean limit,
couplings beyond nearest neighbour have significant weight.
This is demonstrated by the difference between the blue and
red curves, for the NN coupling and the sum of all couplings,
respectively. However when $U/t \gg 1$ the nearest neighbour
coupling dominates. 

%-----------------------------------------------------------------
\begin{figure}[b]
\centerline{
\includegraphics[width=4.2cm,height=4.2cm]{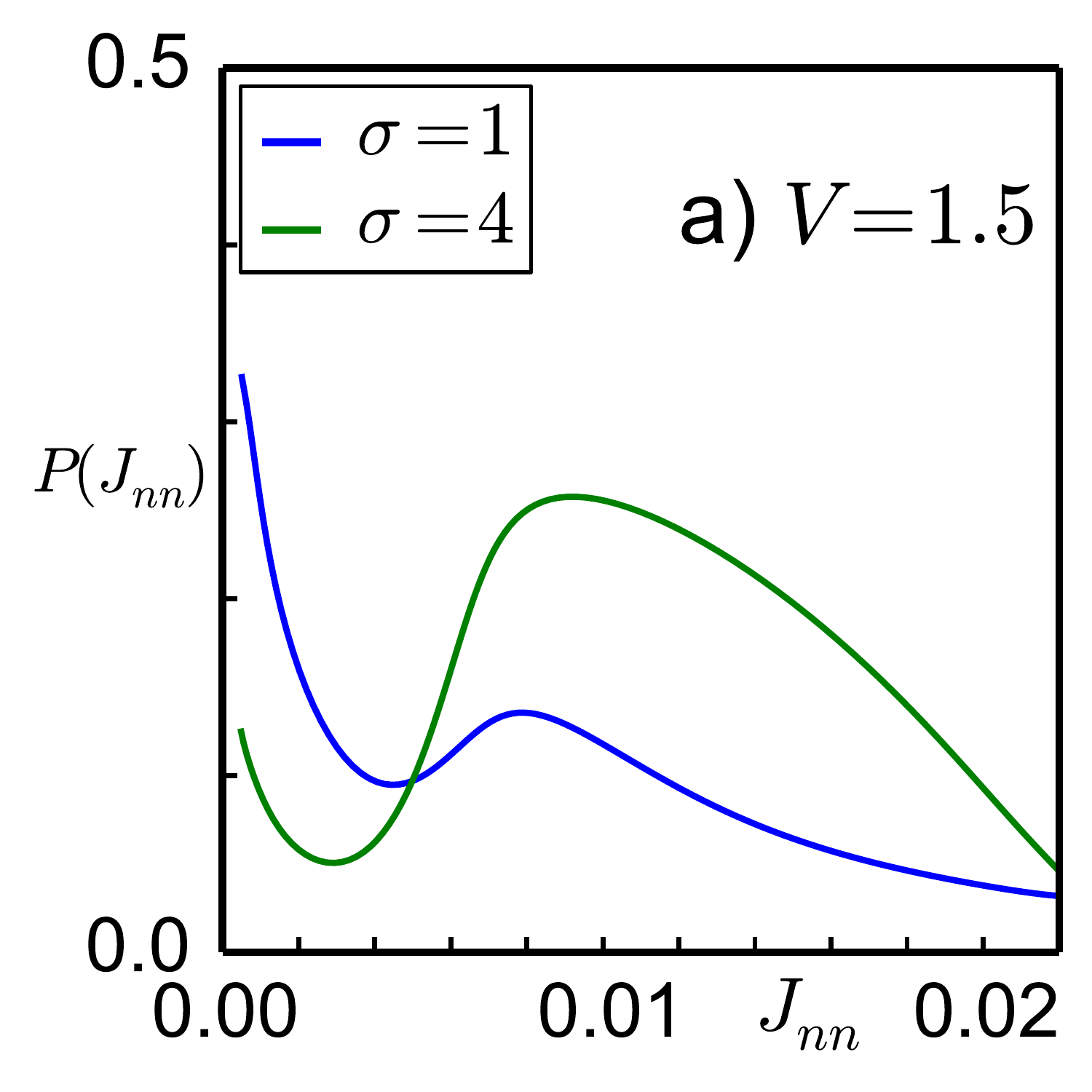}
\includegraphics[width=4.2cm,height=4.2cm]{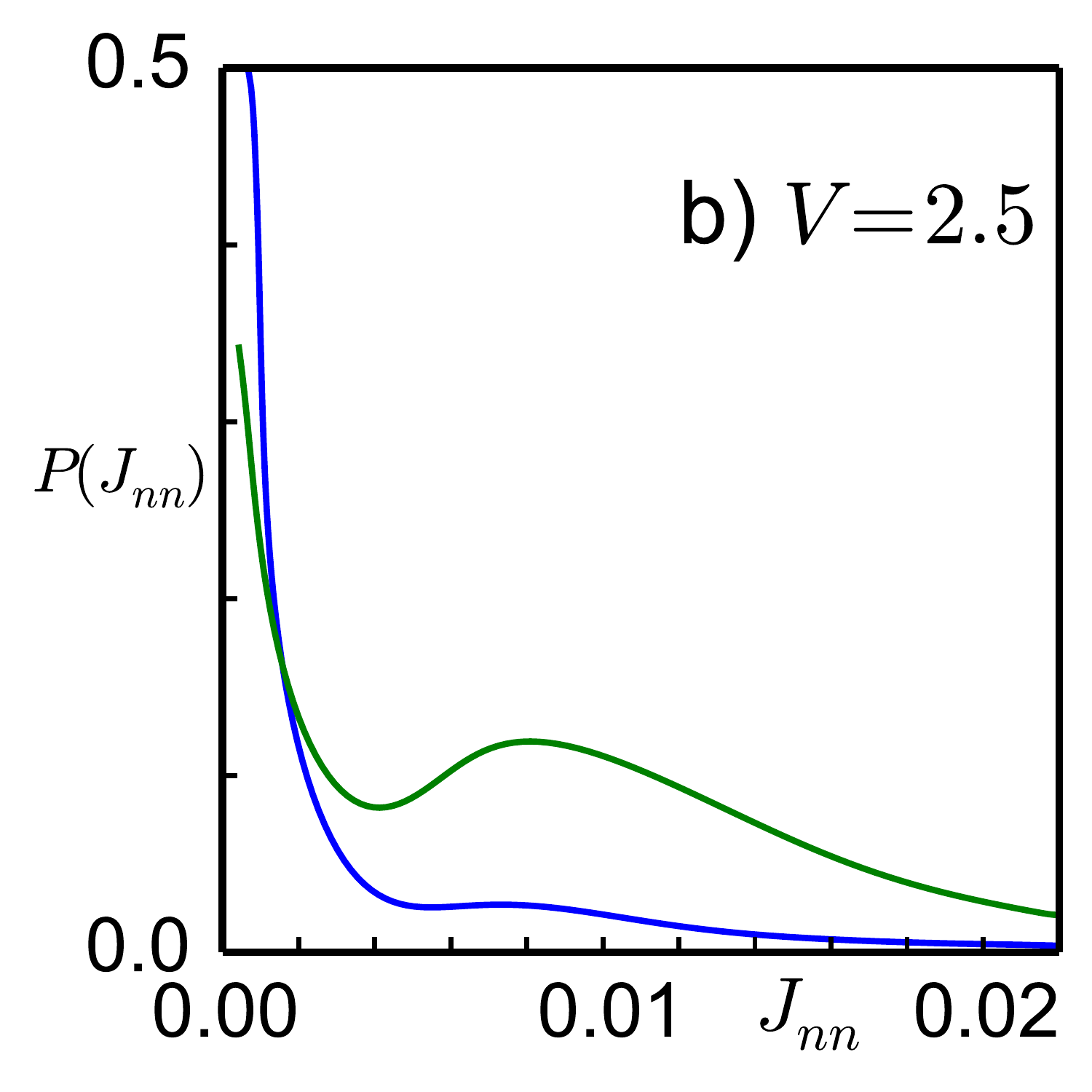}
}
\caption{Distribution of nearest neighbour bonds, averaged over
the system and disorder configurations (a) $\sigma=1$ and (b)
$\sigma=4$.  With increase in disorder strength distribution becomes
broad and peak of the distribution shifts to small values of $J_{nn}$.
At the same disorder the peak is $P(J_{nn})$ is at a larger $J_{nn}$
at larger $\sigma$. This is the origin of the larger $T_c$.
}
\end{figure}
%-----------------------------------------------------------------

At small $U$, one can drop the dependence of the $J_{ij}^0$ 
on the $\Delta_{i0}$ so the overall 
$J_{ij} \propto \Delta^2$. This vanishes as $U/t \rightarrow 0$.
At large $U$, $J_{ij}^0 \sim {1\over \Delta^3}$
so $J_{ij} \sim {1\over \Delta} \sim 1/U$ capturing the large
$ U$ asymptote. The model interpolates between the small $U$
and large $U$ limits. 

Now disorder.  We use the approach above to compute $J_{ij}$.
In presence of disorder effective $J_{ij}$ become
inhomogeneous.  Fig.14 compares the results
of the full MC with that of $H_{XY}$ for 
two speckle sizes. There is a discrepancy 
at small disorder, traceable to the clean 
results at intermediate $U$,  Fig.13 (we are working at
$U=4t$), but the match improves at large disorder.

To understand the effect of disorder on the $J_{ij}$ 
we plot the
distribution of  nearest neighbour bond  $J_{nn}$ for speckle
sizes $\sigma=1,~4$ in Fig.15. 
At both $V=1.5t$ and $2.5t$ the $P(J)$ has a strong peak
at low $J$ when $\sigma $ is small. By contrast most of
the weight at large $\sigma$ is concentrated at intermediate
$J$. The $P(J)$ forms an interesting counterpoint to
the $P(\vert \Delta \vert)$ that we have seen before.
The large $\sigma$ system has relatively smaller number of
large $\Delta_i$ sites coupled strongly and - referring to
the spatial maps - in a  percolative pattern.

\textcolor{red}{
\subsection{Connection to cold atomic experiments}
}

\textcolor{red}{
To the extent we know, superfluidity of fermions in an {\it
optical lattice} has not been observed yet, although superfluidity
in the `continuum', {\it i.e}, in a trap has been achieved.
This is related to the lower $T_c$ (and entropy level) needed
to achieve lattice superfluidity. Given this, there are no
experiments yet that test out the effect of disorder, speckle or
otherwise, on lattice Fermi superfluids. However, most of
the qualitative features that we observe on increasing
speckle size, {\it e.g}, the
increase in $T_c$, the increase in low energy spectral weight,
and the weakening localisation, are not lattice specific features.
These effects should be visible in the continuum case as well.
}

\textcolor{red}{
Specifically, (i)~The change in $T_c$ on increasing speckle size 
can be studied by tracking the condensate fraction via time of
flight measurements. Such measurements are standard in
clean superfluids and  can be  adapted to the disordered case  
\cite{5,6,7,8}.
(ii)~The suppression of the gap in the global DOS on increasing
speckle size  can be probed via radio frequency (RF) spectroscopy,
already used in several cold atom experiments \cite{9,10,12}. 
In fact there is
now a proposal to measure the {\it local} DOS via an
``energy-resolved atomic scanning probe'' \cite{16}. If such a method
is implemented it would directly visualise the order parameter
variation across the speckle disordered sample.
(iii)~Localisation effects in the disordered potential, and their
weakening with increasing speckle size, can be probed via
`expansion' of the disordered gas on removing the trapping
potential\cite{exp-fermi-non-int0,exp-fermi-non-int1}.
(iv)~We did not consider transport effects since our degrees
of freedom were supposed to be neutral (atoms). However, mass
transport measurements in such disordered superfluids 
are already possible \cite{exp-fermi-int2}, 
and the impact of increasing speckle size at
fixed disorder would be fascinating to observe.
}

\section{Conclusion}

We studied the speckle disorder driven superfluid-insulator
transition for intermediate coupling fermions in a 
two dimensional lattice.
The speckle disorder has an exponential on site
distribution, and a correlation length $\sigma$. We observe
the increase of the superfluid window at $T=0$, as well as
increase in $T_c$, with increasing speckle size.
In contrast to the disorder driven superfluid to 
insulator transition, which  is well studied,  we  mapped 
out a speckle size driven {\it insulator to superfluid} 
transition.  While some of the effects of increasing speckle
size are crudely like a decrease in disorder, the underlying
physics is more complex and contradicts this naive expectation.

Growing speckle size at strong disorder 
leads to an energy landscape with large scale undulations. 
In such a background the pairing amplitude is large only
in a small fraction of sites. The small amplitude on the
rest of the sites leads to suppression of the overall spectral
gap, unlike what one would expect from an effective
decrease in disorder. 
The smooth background leads to greater delocalisation of
single particle states which generate an 
effective intersite coupling that grows with speckle size. 
This compensates for having fewer sites having a large 
$\Delta_i$ and leads to a higher $T_c$.
The variation in the superfluid window with speckle size,
the increase in $T_c$, and the unusual low energy spectral
feature, are testable predictions from our work.

{\it Acknowledgements:} 
We acknowledge use of the HPC  Clusters at HRI and
thank Sauri Bhattacharyya for a reading of the manuscript.  
AJ thanks an Infosys grant for support.

\appendix

\section{Derivation of the effective XY model}

We outline here the derivation of effective model.
The partition function of the fermion-auxiliary field model 
is given by
\begin{eqnarray}
Z&=&\int D[\Delta,\Delta^*]D[\phi]D[\psi,\overline{\psi}]e^{-S_{eff}}\cr
S_{eff}&=&\int_0^\beta d\tau 
[ \sum\limits_{i,\sigma} \overline{\psi}_{i,\sigma}{\partial_\tau}
\psi_{i,\sigma}+H_{eff}] \cr
H_{eff} ~&=& H_0 + H_{coup} + 
{1\over \vert U \vert }
\sum\limits_i (|\Delta_i|^2+\phi_i^2) \cr
H_0~~~ &=& -t\sum_{<ij>}^{\sigma} {\bar \psi}_{i\sigma} \psi_{j\sigma}
+\sum\limits_{i\sigma}(V_{i}-{\mu}) {\bar \psi}_{i\sigma} {\psi}_{i\sigma} \cr
H_{coup} & = & ~~~~ \sum_{i} (\Delta_i {\bar \psi}_{i\uparrow} 
{\bar \psi}_{i\downarrow} 
+ h.c) 
+ \sum_i \phi_i \rho_i \nonumber
\end{eqnarray}

Now we approximate
$\phi_i=\phi_{0,i}$
where $\phi_{0i}$ is the $T=0$ saddle point 
value of $\phi$ field and 
similarly we approximate $\Delta_{i}=\Delta_{0i}+\delta\Delta_{i}$
where $\Delta_{0i}$ is the $T=0$ saddle point value of $\Delta$
field.  $S_{eff}$ can be rewritten as
\begin{eqnarray}
S_{eff}&\approx&S_1+S_2+S_3+S_4\cr
S_{1}&=&\int_0^\beta d\tau \sum\limits_{i,\sigma}
\overline{\psi}_{i,\sigma}({\partial_\tau}-\mu)
\psi_{i,\sigma}+\sum_{i} (\Delta_{0i}\overline{\psi}_{i\uparrow}
\overline{\psi}_{i\downarrow} + h.c)\cr
&+&\sum_i \phi_{0i} n_i+H_{0i} + 
{1\over \vert U \vert }
\sum\limits_i (|\Delta_{0i}|^2+\phi_{0i}^2)\cr
S_2&=&\int_0^\beta\sum_{i} (\delta\Delta_{i}\overline{\psi}_{i\uparrow}
\overline{\psi}_{i\downarrow} + h.c)\cr
S_3&=&{\beta\over \vert U \vert }
\sum\limits_i (\delta\Delta^*_i\Delta_{0i}+\delta\Delta_i\Delta_{0i}^*)\cr
S_4&=&{\beta\over \vert U \vert }\sum\limits_i\delta\Delta^*_i\delta\Delta_i\nonumber
\end{eqnarray}

The original partition function is approximated by
\begin{eqnarray}
&&Z\approx\int D[\delta\Delta,\delta\Delta^*]
D[{\psi},\overline{{\psi}}]e^{-S_{1}-S_3-S_4}(\sum\limits_n{(-S_2)^n\over n!})\cr
&=&\int D[\delta\Delta,\delta\Delta^*]
D[{\psi},\overline{{\psi}}]e^{-S_{1}-S_3-S_4}(1-S_2+{(S_2)^2\over 2!}+...)\cr
&=&\int D[\delta\Delta,\delta\Delta^*]Tr[e^{-\beta H_f}]
e^{-S_3-S_4}(1-\braket{S_2}+{\braket{S_2^2}\over 2!}+...)\nonumber
\end{eqnarray}

where 
\vspace{0.35cm}

\hspace{0.7cm}$
\int D[c,\overline{c}]e^{-S_{1}}=Tr[e^{-\beta H_f}]\nonumber
$
\begin{eqnarray}
H_f &=& H_0+\sum_{i} ((\Delta_{0i}c^\dagger_{i\uparrow}
c^\dagger_{i\downarrow} + h.c) 
+ \phi_{0i} n_i )  \cr
~~~~~~~~~~ && + {1\over \vert U \vert }
\sum\limits_i (|\Delta_{0i}|^2+\phi_{0i}^2)\nonumber
\end{eqnarray}
Now
\begin{eqnarray}
\int D[c,\overline{c}]e^{-S_{1}}S_2 &=&  \braket{S_2} \cr
 S_3-\braket{S_2} &=&0 \nonumber
\end{eqnarray}
$$
Z\approx\int D[\delta\Delta,\delta\Delta^*]Tr[e^{-\beta H_f}]
e^{-S_4}(1+{\braket{S_2^2}-\braket{S_2}^2\over 2!}+...)
$$ 
\begin{eqnarray}
{\braket{S_2^2}-\braket{S_2}^2\over 2!}
={1\over 2!}\int_0^\beta d\tau_1 d\tau_2
\sum_{ij}X_{ij}(\tau_1,\tau_2)\nonumber
\end{eqnarray}
\begin{eqnarray}
X_{ij}(\tau_1,\tau_2)&=&-\delta\Delta_i\delta\Delta_j F_{ij}(\tau_1-\tau_2)
F_{ji}(\tau_2-\tau_1)\cr
&&-\delta\Delta_i^*\delta\Delta_j^* F_{ji}
(\tau_1-\tau_2)F_{ij}(\tau_2-\tau_1)\cr
&&+\delta\Delta_i^*\delta\Delta_j G_{ij\uparrow}
(\tau_1-\tau_2)G_{ij\downarrow}(\tau_1-\tau_2)\cr
&&+\delta\Delta_i\delta\Delta_j^* G_{ji\downarrow}
(\tau_2-\tau_1)G_{ji\uparrow}(\tau_2-\tau_1)\nonumber
\end{eqnarray}

$F_{ij}(\tau_2-\tau_1))=Tr[e^{-\beta H_{eff}}T_{\tau}c^\dagger_{i\uparrow}
(\tau_2)c^\dagger_{j\downarrow}(\tau_1)]/Tr[e^{-\beta H_{eff}}]$\\

$G_{ij\sigma}(\tau_2-\tau_1))=Tr[e^{-\beta H_{eff}}T_{\tau}c_{i\sigma}
(\tau_2)c^\dagger_{j\sigma}(\tau_1)]/Tr[e^{-\beta H_{eff}}]$\\

Let's assume small angular fluctuations
$\delta\Delta_i=\iota \theta \Delta_{0i} $

\vspace{0.5cm}
${\braket{S_2^2}-\braket{S_2}^2\over 2!}
=\beta\sum\limits_{ij}\theta_i\theta_j \Delta_{0i}\Delta_{0j}J_{ij}$

\vspace{0.5cm}
$J_{ij}^0={1\over\beta}\sum\limits_{n}
[F_{ij}(\iota\omega_n)F_{ji}(\iota\omega_n)
+G_{ij\uparrow}(\iota\omega_n)G_{ij\downarrow}(-\iota\omega_n)]
$ 

\begin{eqnarray}
J_{ij}^0&=&\sum\limits_{n1,n2}{(u_{n1}^{i*}v_{n1}^{j}u_{n2}^{i*}v_{n2}^{j}
+v_{n1}^{i}u_{n1}^{j*}v_{n2}^{i}u_{n2}^{j*}\over E_{n1}+E_{n2}}\cr
&&+{u_{n1}^{i}u_{n1}^{j*}u_{n2}^{i}u_{n2}^{j*}
+v_{n1}^{i*}v_{n1}^{j}v_{n2}^{i*}v_{n2}^{j}\over E_{n1}+E_{n2}})\nonumber
\end{eqnarray}

where $E_{n1}$ and $E_{n2}$ are eigenvalues and
$u_n$ and $v_n$ are eigenvectors of HBDG problem in presence of
$\Delta_{0i}$ and $\phi_{0i}$.

\vspace{0.5cm}
${\braket{S_2^2}-\braket{S_2}^2\over 2!}\approx-\beta\sum\limits_{ij}
 \Delta_{0i}\Delta_{0j}J_{ij}^0(1-\cos({\theta_i-\theta_j}))$

$$Z_{approx}\approx Tr[e^{-\beta H_f}]
e^{-\beta\sum\limits_{ij} \Delta_{0i}\Delta_{0j}J_{ij}^0(1-\cos({\theta_i-\theta_j}))}$$
So 
the effective model for phase fluctuations over the ground state has the form:
$$
H_{XY}=
\sum\limits_{ij} \Delta_{0i}\Delta_{0j}J_{ij}^0(1-\cos({\theta_i-\theta_j}))
$$ 
Subtracting a constant, this is the form used in the text.


\begin{thebibliography}{99}
\bibitem{footnote} Strictly speaking there is 
no superfluid-insulator
transition at $T=0$ within the HFBdG scheme since quantum
phase fluctuations are absent. However, at any small finite
$T$ thermal fluctuations can cause a SIT.
\end{thebibliography}
\end{document}